\documentclass[aps,twocolumn,amsmath,amssymb,10pt,floatfix,pre]{revtex4-1}
\usepackage{graphicx,float}

\newcommand{\ket}[1]{|#1\rangle}
\newcommand{\pr}{\text{pr}}

\newcommand{\Exp}[1]{\text{e}^{ #1 } }

\begin{document}

\title{Quantum Hamiltonian daemons: unitary analogs of combustion engines}

\author{Eike P.~Thesing, Lukas Gilz and James R.~Anglin}
\affiliation{State Research CENTER OPTIMAS and Fachbereich Physik, Technische Universit\"at Kaiserslautern,\\Erwin Schr\"odinger Stra\ss e 46\\D-67663 Kaiserslautern\\
Germany}

\date{\today}

\begin{abstract}
\textit{Hamiltonian daemons} have recently been defined classically as small, closed Hamiltonian systems which can exhibit secular energy transfer from high-frequency to low-frequency degrees of freedom (\textit{steady downconversion}), analogous to the steady transfer of energy in a combustion engine from the high Terahertz frequencies of molecular excitations to the low kilohertz frequencies of piston motion \cite{Us_PRE}. Classical daemons achieve downconversion within a small, closed system by exploiting nonlinear resonances; the adiabatic theorem permits their operation but imposes non-trivial limitations on their efficiency. Here we investigate a simple example of a quantum mechanical daemon. In the correspondence regime it obeys similar efficiency limits to its classical counterparts, but in the strongly quantum mechanical regime the daemon operates in an entirely different manner. It maintains an engine-like behavior in a distinctly quantum mechanical form: a weight is lifted at a steady average speed through a long sequence of quantum jumps in momentum, at each of which a quantum of fuel is consumed. The quantum daemon can cease downconversion at any time through non-adiabatic Landau-Zener transitions, and continuing operation of the quantum daemon is associated with steadily growing entanglement between fast and slow degrees of freedom.\\
\end{abstract}

\maketitle
\section{Introduction}\label{sec:intro}

From cells to spacecraft, many important dynamical systems convert chemical energy into motion. The short time scales of molecular excitations ($\sim 10^{-14}$s) provide high energy within small mass and volume, but bringing this energy down to much lower frequencies normally proceeds in long cascades, within large dynamical systems that must be described in statistical terms. In dynamical systems with fewer degrees of freedom, secular transfer of energy across a large frequency gap (\textit{steady downconversion}) is usually prevented by \textit{adiabatic decoupling}---the tendency of a rapid subsystem to effectively renormalize the Hamiltonian of a slow subsystem to which it is coupled, but not steadily give the slow system energy.

It has recently been shown, however, that steady downconversion is possible classically within small, closed dynamical systems (\textit{Hamiltonian daemons}) that feature certain highly nonlinear couplings \cite{Us_PRE}. Here we examine the quantum mechanical behavior of such a daemon, using a slightly simplified version of the same example system presented in \cite{Us_PRE}. We find that steady downconversion persists even in quantum regimes within which the essential mechanism of classical daemons can no longer operate, but with specifically quantum mechanical features, including the steady growth of entanglement between fast and slow subsystems.

Our paper is organized as follows. We briefly review the properties of a simple classical daemon as introduced in \cite{Us_PRE}, in a phase space representation that has a natural quantum analog. We discuss the semi-classical limit in which quantum daemons are essentially classical daemons with the addition of Bohr-Sommerfeld quantization, and then focus mainly on the extremely non-classical limit in which the classically crucial resonant region in phase space is too small to support many Bohr-Sommerfeld states---or even too small to support any at all. We show that engine-like behavior persists nonetheless, and discuss the implications of this distinctively quantum mechanical daemon for possible microscopic extensions of thermodynamics.

\section{A simple classical daemon}\label{sec:setup}
\subsection{The daemon Hamiltonian}
After certain approximations described in \cite{Us_PRE}, the time-independent Hamiltonian studied in \cite{Us_PRE} can be represented as
\begin{align}\label{HC}
H&=\frac{P^2}{2M}+ Mg Q + \Omega L_{z}\nonumber\\
&\qquad - \gamma [L_{x}\cos(kQ)+L_{y}\sin(kQ)]\;,
\end{align}
where $Q$, $P$ and $M$ are the height, vertical momentum, and mass, respectively, of a weight that will be raised against gravity $g$ by downconversion. The angular momentum $\mathbf{L}$ represents the fast degree of freedom, with high natural frequency $\Omega$. The rate $\gamma$ and inverse length $k$ are coupling parameters. An explanation of the relationship between \eqref{HC} and the model of Ref.~\cite{Us_PRE} is given in Appendix A.
 
The length $L$ of $\mathbf{L}$ is a constant of the motion under \eqref{HC}, and it sets a convenient action scale. We can use it to represent $\mathbf{L}$ with a single canonical variable pair $(\phi,L_{z})$, by defining $L_{x}=\sqrt{L^{2}-L_{z}^{2}}\cos\phi$ and $L_{y}=\sqrt{L^{2}-L_{z}^{2}}\sin\phi$. This satisfies the canonical Poisson brackets for angular momentum, and yields $L_{x}\cos(kQ)+L_{y}\sin(kQ)=\sqrt{L^{2}-L_{z}^{2}}\cos(kQ-\phi)$. The square root remains real because evolution under \eqref{HC} maintains $|L_{z}|\leq L$ at all times if it is true initially. 

For numerical treatment we can then convert \eqref{HC} to dimensionless variables based on $L$ and $k$: $\tilde{\mathbf{L}}=\mathbf{L}/L$, $\tilde{Q}=kQ$, $\tilde{P} =P/(kL)$. If we also define the dimensionless time $\tau = Mgt/(kL)$, then the canonical equations of motion for evolution in $t$ under \eqref{HC} are equivalent to those obtained for evolution in $\tau$ by treating the $(\tilde{Q},\tilde{P})$ and $(\phi,\tilde{L}_{z})$ as canonical pairs in the dimensionless Hamiltonian
\begin{align}\label{HCtilde}
\tilde{H}&=\frac{\tilde{P}^2}{2\tilde{M}}+ \tilde{Q} + \tilde{\Omega} \tilde{L}_{z} - \tilde{\gamma}\sqrt{1-\tilde{L}_{z}^{2}}\cos(\tilde{Q}-\phi)\;,
\end{align}
with the dimensionless co-efficients
\begin{align}\label{coeffs}
\tilde{M} = \frac{M^{2}g}{k^{3}L^{2}}\qquad \tilde{\Omega}=\frac{kL\Omega}{Mg}\qquad\tilde{\gamma}=\frac{k L\gamma}{Mg}\;.
\end{align}

Because quantization introduces the additional action scale $\hbar$ which must be correctly compared to $L$, we will retain the dimensionful variables throughout this paper, but we will numerically compute and plot the dimensionless forms. Furthermore, we will isolate the effects of quantization itself by comparing models with different values of $L/\hbar$, but identical values of the dimensionless coefficients $\tilde{M}$, $\tilde{\Omega}$, and $\tilde{\gamma}$. All of these models would be the same in the classical limit and so their differences will represent purely quantum effects. Which parameter ranges are most interesting will be apparent after reviewing the kinds of classical time evolution which may be generated by \eqref{HC}.
 
\subsection{Classical daemon evolution}
\begin{figure}
\centering
\includegraphics[width=0.475\textwidth]{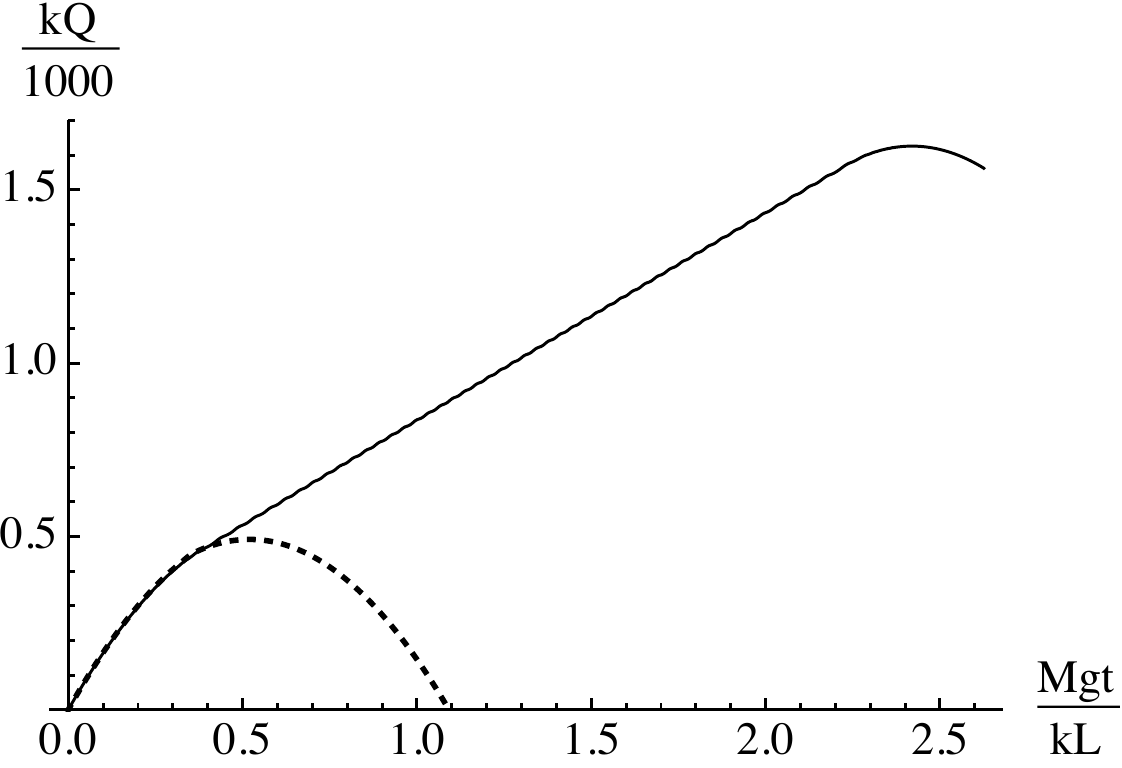}
\caption{\label{fig:ClassEvol1} Illustrative trajectories for the classical daemon with $H$ as given in \eqref{HC}. Trajectories of two types can occur, depending on initial conditions; the sets of initial conditions leading to both represent finite fractions of phase space. The dotted curve shows a trajectory in which downconversion does not take place: the weight is initially launched upwards, and gravity pulls it down in the familiar parabola. The solid curve show a trajectory with downconversion: once it decelerates to $\dot{Q}=v_{c}=\Omega/k$, the weight maintains a very nearly steady upward speed $v_{c}$, with the power needed to rise against gravity being supplied from the high-frequency part of $H$, namely $\Omega L_{z}$. Somewhat before all ostensibly available high-frequency energy has been used (\textit{i.e.} before $L_{z}\to -L$), downconversion ceases and ballistic motion resumes, as discussed in \cite{Us_PRE}. The dimensionless parameters defined in \eqref{coeffs} are chosen here to be $\tilde{M}=1/3000$, $\tilde{\Omega}=600$, and $\tilde{\gamma}=15$, for both trajectories. The initial conditions are $P(0)=0.6 kL$, $kQ(0)=0$, and $L_{z}(0)=\sqrt{5/6}L$, for both trajectories. Only the initial angles are different: $\phi(0)=0$ for solid, $\phi(0)=\pi/2$ for dotted. High numerical precision (25 digits) may be needed to reproduce these exact trajectories from the stated conditions, but qualitatively similar trajectories of both kinds appear without any fine tuning.}
\end{figure}

\begin{figure}
\centering
\includegraphics[width=0.475\textwidth]{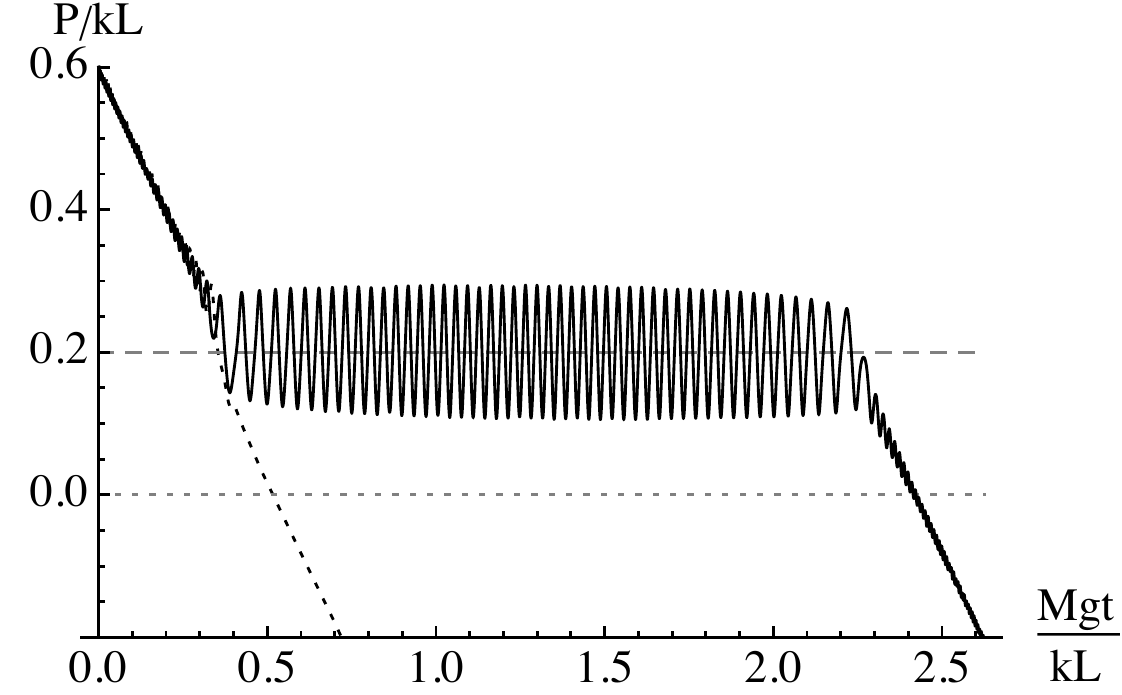}
\caption{\label{fig:ClassEvol2} Momentum $P(t)$ for the same illustrative trajectories shown in Fig.~1. In the decoupling trajectory (dotted), the nonlinear resonance at $P=M v_{c}= \tilde{M}\tilde{\Omega}kL=0.2kL$ produces only a brief perturbation in the steady decline of $P$ due to gravity. The downconversion phase, in contrast, shows steady rapid oscillations around $P=\tilde{M}\tilde{\Omega} kL$, which are also reflected in the slight ripple in the upward slope of the solid curve in Fig.~1. These oscillations may be compared to the cyclic operation of a combustion engine.}
\end{figure}

Using the dimensionless representation just given, we have numerically integrated the canonical equations of motion associated with the Hamiltonian \eqref{HC} for two different representative sets of initial conditions, both with the parameters $\tilde{M}=1/3000$, $\tilde{\Omega}=600$, and $\tilde{\gamma}=15$. The results for $Q(t)$ and $P(t)$, respectively, are shown in Figs.~\ref{fig:ClassEvol1} and \ref{fig:ClassEvol2}. These Figures may be compared to Fig.~1 in Ref.~\cite{Us_PRE}, except that here two different evolutions are shown, for the two different initial conditions.

The two trajectories shown with solid and dashed curves in Figs.~1 and 2 represent the two types of trajectories that can occur under the daemon Hamiltonian \eqref{HC}. The sets of initial conditions leading to both types of trajectory each have infinite measure and represent finite fractions of phase space; the two different $\phi(0)$ values used in Fig.~1 were simply our first two guesses. The dashed curves in Figs.~1 and 2 show a trajectory in which downconversion does not take place: the weight is initially launched upwards, but gravity slows its rise and it eventually falls. The weight's height follows the usual parabola, and its momentum decreases linearly. The effect of the coupling term proportional to $\gamma$ is negligible except for a slight perturbation noticeable near a particular momentum $P=M\Omega/k$ (corresponding in the case shown to $P/(kL)=\tilde{M}\tilde{\Omega}=0.2$). The reason that the coupling makes so little difference in general is \textit{adiabatic decoupling}: because $\Omega$ is large, $L_{x}$ and $L_{y}$ both oscillate rapidly, so that their long term effects average to zero.


In trajectories like the one shown with solid curves in Figs.~1 and 2, on the other hand, what occurs at the special momentum $P=M\Omega/k$ is not a brief perturbation, but a transition to a dramatically different dynamical phase, based on the narrow nonlinear resonance that occurs when the weight rises at the critical speed $v_{c}=\Omega/k$, so that the cosine argument $(\tilde{Q}-\phi)$ in \eqref{HCtilde} becomes slow. The special feature of the daemon is that the resonant effects near $v_{c}$ are able to keep the weight's speed near $v_{c}$, potentially for a long time, even though this means that the weight is steadily rising against gravity. Since total energy is conserved under this time-independent Hamiltonian, the power needed to drive the weight against gravity is supplied from the high-frequency sector's energy $\Omega L_{z}$, which falls steadily. This steady transfer of energy from a high-frequency sector into secular work is \textit{steady downconversion}, and we refer to this dynamical phase of the system as the \textit{downconversion phase}. It cannot continue forever, because $\Omega L_{z}$ is bounded from below for fixed $L$, and so eventually there is a second dynamical transition and the trajectory resumes the usual downward gravitational acceleration (see Ref.~\cite{Us_PRE} for more discussion). We will refer to this simpler type of motion as the \textit{decoupling phase}. The previously discussed type of trajectory, shown in the dashed curves in Figs.~1 and 2, consists entirely of the decoupling phase.

We can now see why certain parameter ranges are of interest. First of all we must have $\tilde{\gamma}\ll\tilde{\Omega}$, because if $\gamma\ll\Omega$ did not hold then the nonlinear coupling would dominate $\Omega L_{z}$ and the basic identification of $\mathbf{L}$ as the high-frequency sub-system would become invalid, leaving a nonlinear dynamical system which might in some ways be interesting but would say nothing about downconversion. Secondly we need $\tilde{\gamma}>1$, since otherwise $d\tilde{P}/d\tau < 0$ always and the coupling can never lift the weight, so only the decoupling phase of motion is possible. Thirdly we are interested in downconversion as a worthwhile power source, such as it is in macroscopic cases like combustion engines and metabolism. The kinetic energy $Mv_{c}^{2}/2$ which must be invested in getting the weight up to the critical velocity should be significantly less than the total high-frequency energy $\sim 2\Omega L$ which is available. This implies the parameter range $\tilde{M}\tilde{\Omega}/4\ll 1$.

All $\tilde{H}$ which fulfill these conditions represent classical Hamiltonian daemons and have qualitatively similar behavior. Presentational goals for this paper, to ensure that both classical and quantum evolutions can be plotted intelligibly on the same axes, impose additional constraints. For example, the dimensionless frequency of the rapid oscillations of $\dot{Q}$ around $v_{c}$ in the downconversion phase can be shown to be of order $\tilde{\omega}=\sqrt{\tilde{\gamma}/\tilde{M}}$; choosing $\tilde{\gamma}$ too large or $\tilde{M}$ too small can easily make these oscillations too rapid to show up well in a plot like Fig.~2. It is important to note, however, that there is in principle no limit to how much work a classical daemon can do, given appropriate parameters. 

\subsection{Adiabatic Decoupling}
The term `adiabatic' refers to situations in which one form or feature of time evolution is much slower than others, such that one may successfully apply approximations based on the smallness of the time scale ratio \cite{Goldstein}. We will refer in this paper, however, to several different kinds of adiabaticity. First of all we have said that the downconversion phase of daemon evolution defies the expectation of \textit{adiabatic decoupling} between systems of inherently different dynamical scales. Adiabatic decoupling is so called because it is a common consequence of multiple time scale evolution. In fact, however, the downconversion phase of our daemon can be described very well within the standard adiabatic approximation. In its downconversion phase the daemon presents an unusual form of adiabatic evolution, because of its particular nonlinear interaction; it violates something that adiabaticity often implies (adiabatic decoupling) but it does not violate adiabaticity itself. 

The changes of dynamical phase from decoupling to downconversion and back, on the other hand, are inherently non-adiabatic processes, which occur through the breakdown of adiabatic approximations in the phase space neighborhood of an unstable fixed point. The change of dynamical phase alters the form of the system's phase space orbits, so that the area enclosed by the orbits---the adiabatically invariant action---not only takes on different values but becomes a completely different quantity. In this sense the full evolution of the classical daemon, including dynamical phase transitions, is necessarily and importantly non-adiabatic. 

Finally, quantum and classical versions of the same Hamiltonian may be adiabatic in different ways or to different degrees, because energy quantization gives the quantum system additional time scales that do not appear in the classical system. As we will observe in Section V, below, the quantum analog of the downconversion phase can in fact begin, continue, and then end, entirely through adiabatic Landau-Zener transitions, even though the classical transitions are necessarily non-adiabatic. This will represent one of several distinct differences between the classical daemon and the strongly quantum mechanical daemon.

\section{The quantum daemon}
\subsection{Quantization}
The system of \eqref{HC} can be quantized by promoting its dynamical variables to operators, denoted by the same variable symbols, but now with circumflex accounts. We impose the canonical commutation relations $[\hat{Q},\hat{P}]=i\hbar$ and $[\hat{L}_{l},\hat{L}_{m}]=i\hbar\sum_{n}\epsilon_{lmn}\hat{L}_{n}$ where $\epsilon_{lmn}$ is the antisymmetric tensor and the values $1,2,3$ of $l,m,n$ are identified with the axes $x,y,z$. Defining $\hat{L}_{\pm}=\hat{L}_{x}\pm i\hat{L}_{y}$ as usual lets us write the quantum analog of \eqref{HC} as
\begin{align}\label{HQ}
\hat{H}&=\frac{\hat{P}^2}{2M}+ Mg \hat{Q} + \Omega\hat{L}_{z} -\frac{\gamma}{2}\left(\hat{L}_{-}e^{ik\hat{Q}}+\hat{L}_{+}e^{-ik\hat{Q}}\right)\;.
\end{align}
The high-frequency energy source $\Omega\hat{L}_{z}$ of the quantum daemon thus has discrete energy levels with uniform spacing $\hbar\Omega$. The effect of the interaction Hamiltonian operator is also easy to see: it either consumes one quantum of `fuel' (lowering $\hat{L}_{z}$ with $\hat{L}_{-}$) and gives the weight an upward momentum kick $\Delta P = \hbar k$, or else restores one quantum of fuel (by acting with $\hat{L}_{+}$) and applies an opposite kick to the weight. 

\subsection{Reduction to the time-dependent effective Hamiltonian}
In Ref.~\cite{Us_PRE} we showed how to exploit a first integral of \eqref{HC} in order to exactly reduce the problem's phase space from four dimensions to two, using a time-dependent canonical transformation which made the effective two-dimensional Hamiltonian time-dependent. While the reduction in \cite{Us_PRE} was to eliminate $\mathbf{L}$ and leave an infinite two-dimensional $(Q,P)$ phase space, it is just as possible classically to eliminate $P$ and $Q$ instead and leave only the compact two-dimensional phase space of $\mathbf{L}$. Here we perform this reduction quantum mechanically, in the Schr\"odinger picture. Although this mapping will be exact in our particular Hamiltonian only because the weight's potential $Mg\hat{Q}$ is linear, a similar mapping will be valid as a Born-Oppenheimer approximation for a wide range of more general models with potentials $V(\hat{Q})$. This is directly analogous to the adiabatic linearization of the classical potential in \cite{Us_PRE}.

We begin with a basis of tensor product states of the slow and fast sectors, of the form $\ket{m}_{f}\ket{P}_{s}$, where by $\ket{P}_{s}$ we mean a continuum-normalized momentum eigenstate of the weight, satisfying $\hat P\ket P_s=P\ket P_s$, while $\ket{m}_{f}$ denotes an eigenstate of $\hat{L}_{z}$ with eigenvalue $m\hbar$. We assume that our state is always an eigenstate of $\hat{L}^{2}=\hat{L}_{z}^{2}+\hat{L}_{x}^{2}+\hat{L}_{y}^{2}$ with eigenvalue $l(l+1)\hbar^{2}$ for $l$ some positive integer or half-integer; this assumption is without loss of generality, since $\hat{L}^{2}$ commutes with $\hat{H}$. The $\hat{L}_{z}$ quantum number $m$ can therefore take values in steps of one from $-l$ to $l$. For comparison with the classical model, including computation of the dimensionless coefficients $\tilde{M}$, $\tilde{\Omega}$ and $\tilde{\gamma}$, we use $L=\sqrt{l(l+1)}\hbar$.

Without loss of generality we may express the total quantum state $\ket{\Psi(t)}$ of our system as
\begin{equation}\label{eq:ansatz1}
\ket{\Psi(t)}= \sum_{m=-l}^l \int^{\infty}_{-\infty}\!dP\,  \Psi_m(P,t) \ket{m}_{f}\ket{P-m\hbar k}_{s}
\end{equation}
for some set of $2l+1$ wave functions $\Psi_{m}(P,t)$. For example, a quantum state $\ket{\Psi^{0}}$ in which the fast sector is in an eigenstate of $\hat{L}_{z}$ with eigenvalue $+l\hbar$ while the weight is in a wave packet of position width $D$, average position $Q_{0}$ and average momentum $P_{0}$ would have
\begin{equation}\label{packet}
\Psi^{0}_{m}(P,t) = \delta_{ml}\,Z\,e^{-\frac{1}{2}D^{2}(P-P_{0}+l\hbar k)^{2}/\hbar^{2}}e^{-iQ_{0}P/\hbar}
\end{equation}
with normalization constant $Z$. 

The Schr\"odinger equation $i\hbar\frac{d}{dt}\ket{\Psi}=\hat{H}\ket{\Psi}$ then implies 
\begin{align}\label{Schr1}
i\frac{\partial}{\partial t}\Psi_{m}(P,t) =& \left[\frac{(P-m\hbar k)^{2}}{2M\hbar}+m\Omega+iMg\frac{\partial}{\partial P}\right]\Psi_{m}(P,t)\\
& - \frac{\gamma}{2}\left[\sqrt{(l-m)(l+m+1)}\Psi_{m+1}(P,t)\right.\nonumber\\
&\qquad+\left.\sqrt{(l+m)(l-m+1)}\Psi_{m-1}(P,t)\right]\nonumber
\end{align}
when we use the standard matrix elements for $\langle m|\hat{L}_{\pm}|n\rangle$ and the identity $e^{i\xi\hat{Q}}|P\rangle = |P+\hbar\xi\rangle$.

By now defining
\begin{align}\label{ReductionMapping}
\Psi_{m}(P,t)&=\psi_{m}\left(P + Mg t,t\right)e^{i\frac{P^{3}}{6M^{2}g\hbar}}
\end{align}
we eliminate the derivative with respect to $P$ from the Schr\"odinger equation for $\psi_{m}(P,t)$, which can be written
\begin{align}\label{E:ReducedSchroedinger}
i\frac{\partial}{\partial t}\psi_{m}(P,t)=& \frac{Mg}{kL} \sum_{n=-l}^{l}h_{mn}(t-t_{P})\psi_{n}(P,t)\nonumber\\
t_P=&\frac{P}{Mg}-\frac{\Omega}{kg}\nonumber\\
h_{mn}(t)=&h_{m}(t)\delta_{mn}-w_{mn}\nonumber\\
h_{m}(t)=&\frac{1}{\tilde{M}}\left(\frac{\hbar}{2L}m^{2}+\frac{Mgt}{kL}\,m\right)\nonumber\\
w_{mn}=&\frac{\tilde{\gamma}}{2}\sum_\pm\sqrt{l\left(l+1\right)-mn}\,\delta_{m,n\pm1}\;.
\end{align}
If we evolve in the dimensionless time $\tau=Mgt/(kL)$ then the coefficients in \eqref{E:ReducedSchroedinger} can all be expressed in terms of the classical dimensionless coefficients $\tilde{M}$, $\tilde{\Omega}$ and $\tilde{\gamma}$, plus the additional ratio $L/\hbar =\sqrt{l(l+1)}$.

Since there is no differentiation with respect to $P$ in \eqref{E:ReducedSchroedinger}, we can solve for $\psi_{m}(P,t)$ just by solving the evolution in the $2l+1$-dimensional Hilbert space of $\psi_{m}$ for $|m|\leq l$ and fixed $P$. The initial conditions on $\psi_{m}(P,t)$ will (in general) depend on $P$, and so does the time offset $t_{P}$; but for each value of $P$, the Schr\"odinger evolution problem only needs to be solved in the reduced Hilbert space of dimension $2l+1$. We can express this in operator notation by defining the reduced state vector 
\begin{equation}
|\psi(P,t)\rangle = \sum_{m=-l}^{l}\psi_{m}(P,t)|m\rangle
\end{equation}
 which depends on $P$ as a c-number parameter. Eqn.~\eqref{E:ReducedSchroedinger} for the $\psi_{m}(P,t)$ is then equivalent to $i\hbar\partial_{t}|\psi(P,t)\rangle = \hat{H}_\text{eff}(t-t_{P})|\psi(P,t)\rangle$ for
\begin{equation}\label{E:qEffectiveH}
	\hat H_\text{eff}(t)=\frac{k^2 \hat L_z^2}{2M}+  g kt \hat L_z-\gamma \hat{L}_x\;.
\end{equation}

The remainder of our analysis will be entirely based on this $\hat H_\text{eff}$. Hamiltonians like \eqref{E:qEffectiveH} have already received significant study and we will not report any unexpected features in the quantum evolution under $\hat H_\text{eff}$. Our contribution in this paper concerns the consequences of this reduced evolution for the operation of the larger system as a quantum daemon, as shown by the quantum mechanical motion of the weight, which is given through the mapping \eqref{ReductionMapping}. For readers who are familiar with previous studies of $\hat H_\text{eff}$, however, we pause here to explain why the features of $\hat H_\text{eff}$ which are important for daemons are essentially orthogonal to those that are interesting from other perspectives. Other readers may wish to skip to the next Section.
%Where by $L_{x,z}$ we mean the operator with operator action
%\begin{align}
	%\hat L_z\ket m&=\hbar m\ket m,\\
	%\hat L_x\ket m&=\hbar\sum_\pm\sqrt{l(l+1)-m(m\pm 1)}\ket{m\pm1},\\
	%&\text{where}\quad e^{-i\hbar\int_0^t\hat H_{\text{eff}}(t')dt' }\langle p \vert m\rangle=\psi_m(p,t).\nonumber
%\end{align}
%We emphasize at this point that due to transformation \eqref{ReductionMapping} $\ket m$ and $\ket m_f$ are not the same basis. The notation $\hat L_z$ and $\hat L_x$ merely implies that both operators act on the new basis in the way we expect angular momentum operators to act.

\subsection{Features of $\hat H_\text{eff}$}
Classical systems with Hamiltonians like \eqref{E:qEffectiveH} have been studied either as nonlinear generalizations of the quantum mechanical problem of two-state Landau-Zener tunneling \cite{Liu}, or as mean-field descriptions of Josephson junctions and trapped Bose-Einstein condensates in tunable double wells. Several papers have analyzed cases with more complicated time dependence, in order to achieve optimal population transfer by external control \cite{Liew,Nesterenko,Dou}. In our Hamiltonian daemon context the strictly linear $g(t-t_{P})$ term is what appears by reduction from our original time-independent Hamiltonian, and so only this simpler case is of interest to us.

Other works have considered models equivalent to \eqref{E:qEffectiveH} simply as instructive examples of adiabatic dynamics in non-linear quantum systems \cite{Liu,Fu,Pu,Trimborn}. In these cases attention has focused mainly on evolution from the initial state $\psi_{m}=\delta_{m,-l}$, as it relates to classical motion  starting at the ``south pole'' of the $\mathbf{L}$-sphere. In the ``strongly nonlinear'' classical regime ($\tilde{M}\tilde{\gamma}<1$ in our notation) this initial fixed point does not migrate up towards the top of the $\mathbf{L}$-sphere, as it does in the weakly nonlinear ($\tilde{M}\tilde{\gamma}>1$) regime. Instead for $\tilde{M}\tilde{\gamma}<1$ the initial fixed point at $L_{z}=-L$ meets an unstable fixed point and disappears, leaving the system at a negative $L_{z}$ at late times instead of near $L_{z}=+L$, in dramatic contrast both to the classical adiabatic behavior with $\tilde{M}\tilde{\gamma}>1$ and to (the extreme adiabatic limit of) the corresponding quantum adiabatic evolution in all cases \cite{Liu,Fu,Trimborn}.  

In contrast to those previous studies, however, our use of \eqref{E:qEffectiveH} as an exact reduced description of the time-independent daemon Hamiltonian makes us interested only in evolution from the top of the $\mathbf{L}$-sphere downward, so that the fast subsystem transfers energy to the weight. This classical evolution is much more conventionally adiabatic, involving orbits around a single instantaneous fixed point that persists and migrates. The subtle issues of disappearing fixed points and the ordering of classical and adiabatic limits are essentially irrelevant for us. Our results concerning \eqref{E:qEffectiveH} itself will be unsurprising. Our contribution is to show what these straightforward results imply for the quantum daemon.

\section{The semi-classical daemon}
\subsection{The classical limit}
For $L/\hbar\to\infty$ one recovers the classical limit of \eqref{E:qEffectiveH}; using the classical canonical variables $L_{z}$ and $\phi$ that we defined for \eqref{HCtilde}, this can be written
\begin{equation}\label{HCred}
	H_\text{eff}(t)=\frac{k^2 L_z^2}{2M}+  g kt L_z-\gamma \sqrt{L^{2}-L_{z}^{2}}\cos(\phi)\;.
\end{equation}
(We show in Appendix B how this classical $H_\text{eff}$ may be obtained directly from the classical $H$ by a time-dependent canonical transformation.)
The rate at which the contours of constant $H_\text{eff}$ change can be recognized for small $\tilde{M}\tilde{\gamma}$ by setting $gktL\sim k^{2}L^{2}/M$ and obtaining the time scale $kL/(Mg)$ which (not coincidentally) has defined our dimensionless time $\tau$. The typical rate at which the classical system orbits around those contours can be estimated at $t=0$ as the frequency of small oscillations around the stable fixed point $L_{x}=L$, which for small $\tilde{M}\tilde{\gamma}$ is $\sqrt{\gamma k^{2}L/M}$. The rate of orbiting divided by the rate of orbit change is thus of order $\sqrt{\tilde{\gamma}/\tilde{M}}$, which is large for all parameter ranges that can represent daemons, and so for daemon cases we can expect \eqref{HCred} to be adiabatic. Except in the vicinity of unstable instantaneous fixed points, the adiabatic theorem \cite{Goldstein} says that the system will closely follow contours of constant instantaneous $H_\text{eff}$. As these contours slowly change, the system will remain on that contour which encloses a constant phase space area.

\begin{figure*}[hbt]
	\includegraphics[width=.95\textwidth]{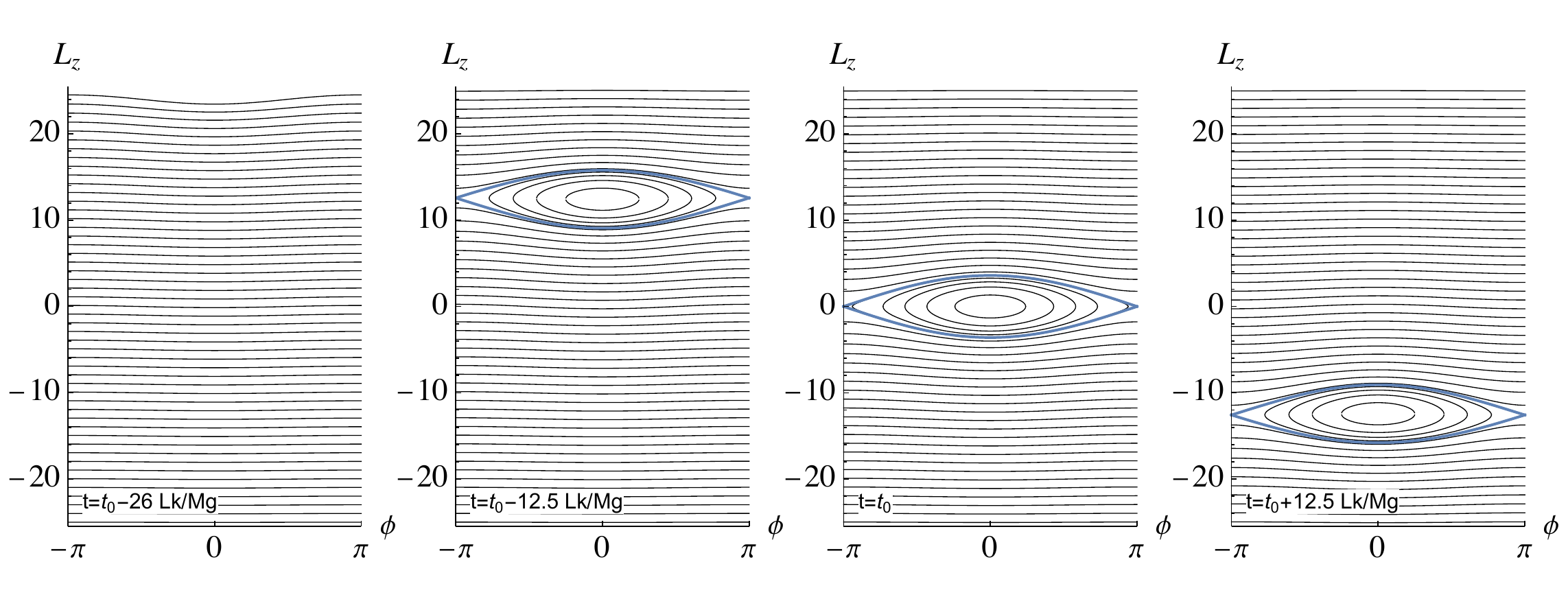}
	\caption{\label{fig:bohr-sommerfeld}Contours of constant instantaneous $H_\text{eff}$ as given by the classical limit \eqref{HCred} of \eqref{E:qEffectiveH}, in the horizontally periodic $(\phi,L_{z})$-plane at different times. Hamiltonian parameters are chosen as in Figs.~\ref{fig:ClassEvol1} and \ref{fig:ClassEvol2}. Thick blue is the separatrix. The other contours are specifically those for semi-classical Bohr-Sommerfeld energy levels with $L=\sqrt{25\cdot 26}\hbar$. At large negative times there is no separatrix, and all instantaneous energy contours wind across $\varphi$ at nearly constant $L_{z}$. As $t$ increases, the separatrix and its basin of downconversion orbits slowly enter  the $(\phi,L_{z})$ plane from above. They descend through it until they eventually exit below; at late times all energy contours will again become horizontal.}
\end{figure*}

To see what this means we can plot the instantaneous energy contours for $H_\text{eff}(t)$, for different $t$, as functions of the classical canonical variables $L_{z}$ and $\phi$. As Fig.~\ref{fig:bohr-sommerfeld} shows, the adiabatic orbits are in general of two different forms, because for a range of fixed values of $t$ there is a separatrix (the thick, blue curves in the Figure), the two points of which meet at an unstable instantaneous fixed point, at some $t$-dependent height along the side borders $\phi=\pm\pi$. As $t$ increases, the separatrix will first appear close to the $L_{z}=+L$ top of the accessible phase space, and then steadily move down towards the $L_{z}=-L$ bottom, where it will eventually vanish again. The first panel of  Fig.~\ref{fig:bohr-sommerfeld} shows the uppermost orbits already slightly deformed before the separatrix first appears. The basin of bound orbits inside this separatrix represents the self-sustaining nonlinear resonance which permits steady downconversion in classical Hamiltonian daemons \cite{Us_PRE}.

Above and below this separatrix when it exists, and otherwise over the entire plane, the contours of constant instantaneous $H_{\mathrm{eff}}$ are deformed horizontals. These orbits conform more and more closely to exact horizontals, \textit{i.e.}, to contours of constant $L_{z}$, the further away they are from the separatrix. Since constant $L_z$ also means constant high-frequency energy, these orbits do not feature downconversion; the high-frequency energy remains adiabatically conserved. These orbits represent the decoupling dynamical phase of the system.

The classical downconversion phase is represented by trajectories in which the system is captured inside the separatrix, adiabatically following an orbit about the stable instantaneous fixed point which is located on the vertical line $\phi=0$. As the separatrix moves towards the bottom of the phase space plane, this instantaneous fixed point and all the adiabatic orbits around it move down as well; if the system is on one of these adiabatic orbits, then it will be adiabatically dragged along with them. Because $\Omega L_z$ is the high-frequency energy, the adiabatic dragging of the system down to lower average values of $L_{z}$ represents steady downconversion. After downconversion has ceased the system orbits along a horizontal in the lower half of the plane; the analysis of \cite{Us_PRE} shows that the final orbit's distance from the bottom $L_{z}=-L$ will be approximately equal (up to post-adiabatic corrections) to the initial orbit's distance from the top $L_{z}=+L$.

\begin{figure}[htb]
	\includegraphics[width=.475\textwidth]{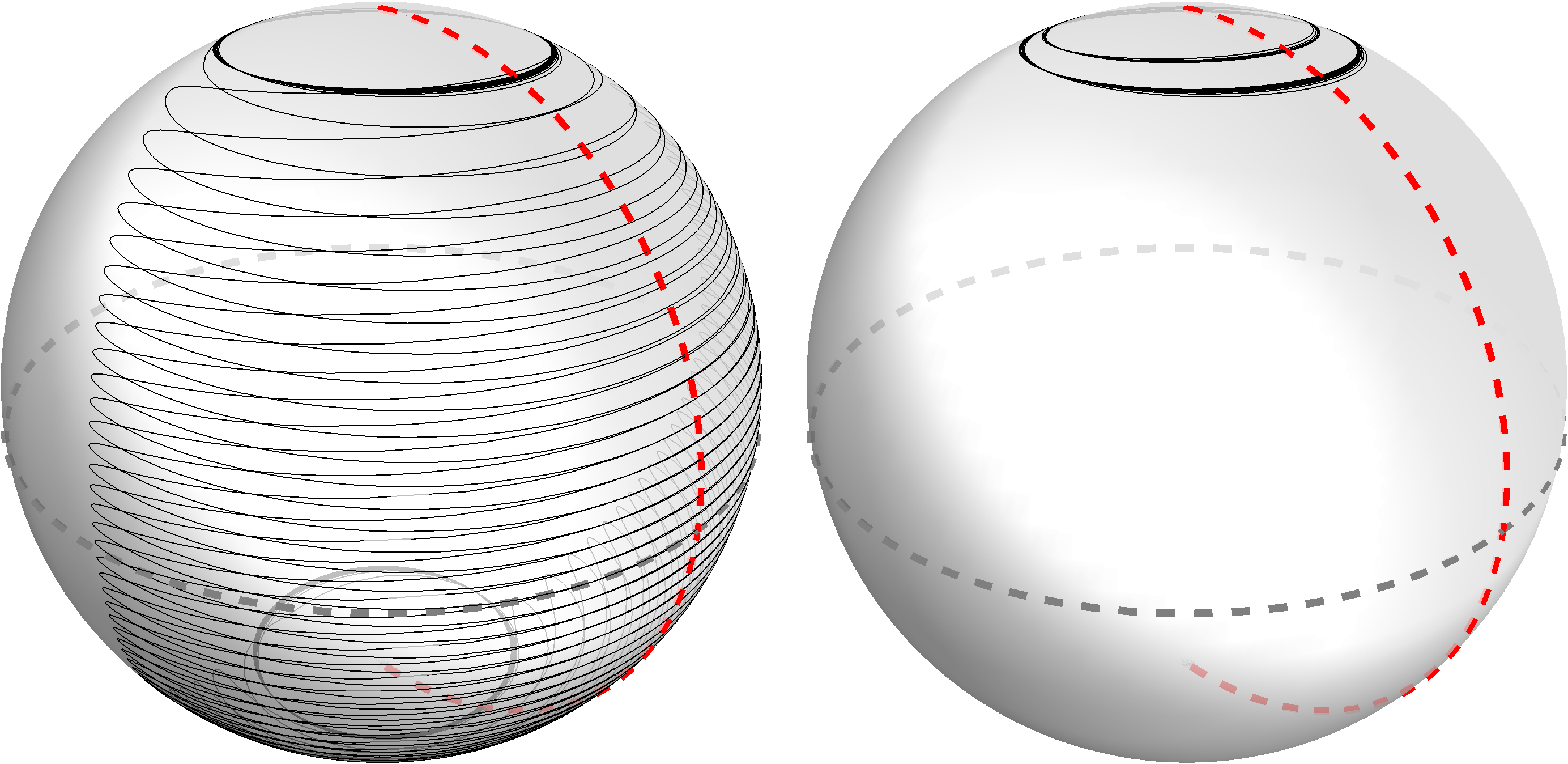}
	\caption{\label{fig:EvolSphere}The evolution on the sphere of $\mathbf{L}$ induced by effective Hamiltonian \eqref{HCred}. Parameters are the same as in Figs.~\ref{fig:ClassEvol1} and \ref{fig:ClassEvol2}; the different orbits shown in left and right plots represent the same two evolutions shown in those Figures (the trajectory that exhibits downconversion is on the left). The dashed gray circle is the equator $L_{z}=0$ and the dashed red arc is the meridian $\phi=0$.}
\end{figure}
The accuracy of the classical adiabatic approximation for this system can be confirmed by exact numerical evolution, as shown in the $\mathbf{L}$-sphere representation in Fig.~\ref{fig:EvolSphere}. The adiabatic approximation breaks down in the vicinity of the unstable fixed point, and so as the slowly moving separatrix approaches the initial orbit, a non-adiabatic transition may occur. The system can either cross into the separatrix (initiating the downconversion phase) or be transferred onto a decoupling orbit of different $L_{z}$ (the daemon `engine' fails to `ignite'). This classical transition is studied elsewhere\cite{ignition_eprint}. In trajectories where the downconversion phase does occur, the downconversion orbits inside the separatrix are eventually expelled again back through the separatrix, because as the separatrix moves downwards through the sphere's lower hemisphere, it steadily shrinks. The point at which this must occur can be derived, up to small post-adiabatic corrections, as being due to the adiabatic invariance of the phase space area enclosed by the downconversion orbits. The explanation of this point in \cite{Us_PRE} can be translated directly into the compact phase space representation that we have constructed here.

\subsection{Bohr-Sommerfeld quantization}
In the limit where $L/\hbar$ is large but not infinite, we may obtain the discrete instantaneous energy levels of the system by using the Bohr-Sommerfeld quantization condition, whereby the discrete allowed energies are those for which the constant-energy contour encloses a phase space area that is a half-integer multiple of $2\pi\hbar$:
\begin{equation}
	\int_c L_z d\phi=2\pi\hbar \left(n+\frac12\right)\;.
\end{equation}
As long as $2\pi\hbar$ is small compared to the total phase space area of the sphere $4\pi L$, we will find a large number of energy levels densely covering the sphere. Except for the separatrix itself, the fifty-one energy contours plotted in Fig.~\ref{fig:bohr-sommerfeld} are actually all for Bohr-Sommerfeld energy levels, plotted for the case $L^{2}=25(25+1)\hbar^{2}$.

Let us consider the evolution of the semiclassical system on one of the energy levels inside the separatrix (\textit{i.e.}, in the downconversion phase). As long as the Hamiltonian changes adiabatically the system will follow the instantaneous eigenstate. In the Bohr-Sommerfeld regime, this means conserving the orbit's enclosed phase space area, exactly as under the classical adiabatic theorem. The only deviation from this behavior will occur when the system's Bohr-Sommerfeld orbit comes close to the classical separatrix. The only significant difference between classical and semiclassical evolution of the daemon, therefore, is that the classical post-adiabatic corrections to the motion near the separatrix must also be supplemented, in the semiclassical regime, by post-Bohr-Sommerfeld quantum corrections. 

These corrections may affect the details of which initial conditions lead, with which probability amplitudes, to a downconversion phase of evolution. They may also provide small corrections to the precise time at which the downconversion phase ends, and hence to the total amount of work that the daemon does by downconversion.  Such corrections must be small in the semiclassical regime, however, simply because the time in which most downconversion orbits are close to the separatrix, and hence show significant quantum corrections, is a small fraction of the total duration of the downconversion phase, during most of which the orbits are well inside the separatrix, and show negligible quantum effects beyond Bohr-Sommerfeld energy quantization itself. In this sense there is qualitatively very little difference between the classical daemon and the quantum daemon in the semiclassical regime. Bohr correspondence of quantum and classical dynamics applies to Hamiltonian daemons.

What happens to quantum daemons, however, when quantum effects become stronger as the action scale of the system, and in particular the phase space area enclosed by the separatrix, becomes less large compared to $\hbar$? We show in Appendix C that for small $\tilde{M}\tilde{\gamma}$ the phase space area enclosed by the separatrix may be estimated as
\begin{equation}
S_\text{sep}\sim 16 L \sqrt{\tilde{M}\tilde{\gamma}}\;.
\end{equation}
The onset of stronger quantum effects on the daemon thus begins when $S_\text{sep}$ is no longer large compared to the Bohr-Sommerfeld action level spacing $2\pi\hbar$. In terms of our dimensionless parameter ratios defined in \eqref{coeffs} this means that
\begin{equation}
\tilde{M}\tilde{\gamma}\lesssim\left(\frac{\pi\hbar}{8L}\right)^{2}
\end{equation}
defines the regime of strong quantum effects.

The intermediate quantum regime, where $\tilde{M}\tilde{\gamma}$ is significantly but not enormously larger than $(\hbar/L)^{2}$, poses interesting problems for future investigation. Here we will focus instead on the strong quantum regime, and ultimately ask what happens to the quantum daemon in the limit where the area enclosed by the separatrix is actually smaller than $\pi\hbar$, so that even the basic idea of having the system captured inside the separatrix becomes incompatible with the Heisenberg uncertainty principle. Everything that we have learned about Hamiltonian daemons from classical mechanics \cite{Us_PRE} tells us that long-term capture of the system inside the separatrix around a localized nonlinear resonance is the essential daemon phenomenon which enables steady downconversion. One might therefore easily imagine that quantum daemons in the extreme quantum limit should simply fail to work at all. We will see instead, however, that even such extreme quantum daemons can still perform steady downconversion very well---in a very quantum mechanical way.

\section{Strongly quantum daemons}
The quantum daemon's dynamics can be described by evolving quantum states under the Hamiltonian (\ref{E:qEffectiveH}) numerically. Here we will consider the strongly quantum regime by examining as an example a case with $l=5$, and otherwise $\tilde{M}=1/3000$, $\tilde{\Omega}=600$ and $\tilde{\gamma}=15$, just as in the classical evolutions shown in Figs.~1, 2 and 4. By considering a quantum initial state with $m=l=5$, so that $L_{z}/L= l/\sqrt{l(l+1)}=\sqrt{5/6}$ initially just as in the classical trajectories of Figs.~1, 2 and 4, we will thus be presenting maximally similar quantum and classical daemon evolutions, differing only in the parameter $L/\hbar$ being $\sqrt{30}$ in the quantum case while it was infinite in the classical cases. Figures 5 and 6 in this Section will therefore directly compare the quantum evolution with that of an initially similar classical ensemble which will be described in sub-section V.B, below.

The parameters for our quantum case here provide $\tilde{M}\tilde{\gamma}=0.005$ and $[(\pi\hbar)/(8L)]^{2}=(\pi/8)^{2}/30\doteq 0.0051$, so our small-$\gamma$ estimate of the separatrix area should be close to $2\pi\hbar$. Numerical integration shows that the actual separatrix area in this case is approximately $1.97\pi\hbar$ when the separatrix is largest, which occurs at the moment when the unstable fixed point is at $(\phi,L_{z})=(\pi,0)$. At the time when classical downconversion begins (if it does), the unstable fixed point would be near $(\phi,L_{z})=(\pi,\sqrt{5/6}L)$ and the separatrix area would be approximately $1.27\pi\hbar$. In this case therefore the separatrix would support exactly one Bohr-Sommerfeld level throughout the evolution. Based on a semi-classical picture of how quantum daemons work, one might therefore expect this quantum daemon to exhibit steady downconversion in just one special state.

We will see that this is indeed what occurs. We will also see, however, that the behavior of the strongly quantum daemon is dramatically non-classical in several ways. Its behavior can be explained quite simply in terms of Landau-Zener transitions at a series of avoided level crossings. This Landau-Zener theory extends straightforwardly, moreover, to arbitrarily small $\tilde{\gamma}\tilde{M}$. It shows clearly that the strongly quantum daemon can behave very much like the case we show in this Section even when the classical separatrix area is much smaller than $\pi\hbar$, so that the classical downconversion mechanism, of system capture within the adiabatic separatrix, becomes impossible. The Landau-Zener mechanism of steady downconversion thus represents an independent mechanism for Hamiltonian daemons which can take over in the strongly quantum regime.

\subsection{Numerical results for the strongly quantum daemon}
\subsubsection{The weight rises---and also does not}
\begin{figure*}
\centering
\includegraphics[width=\textwidth]{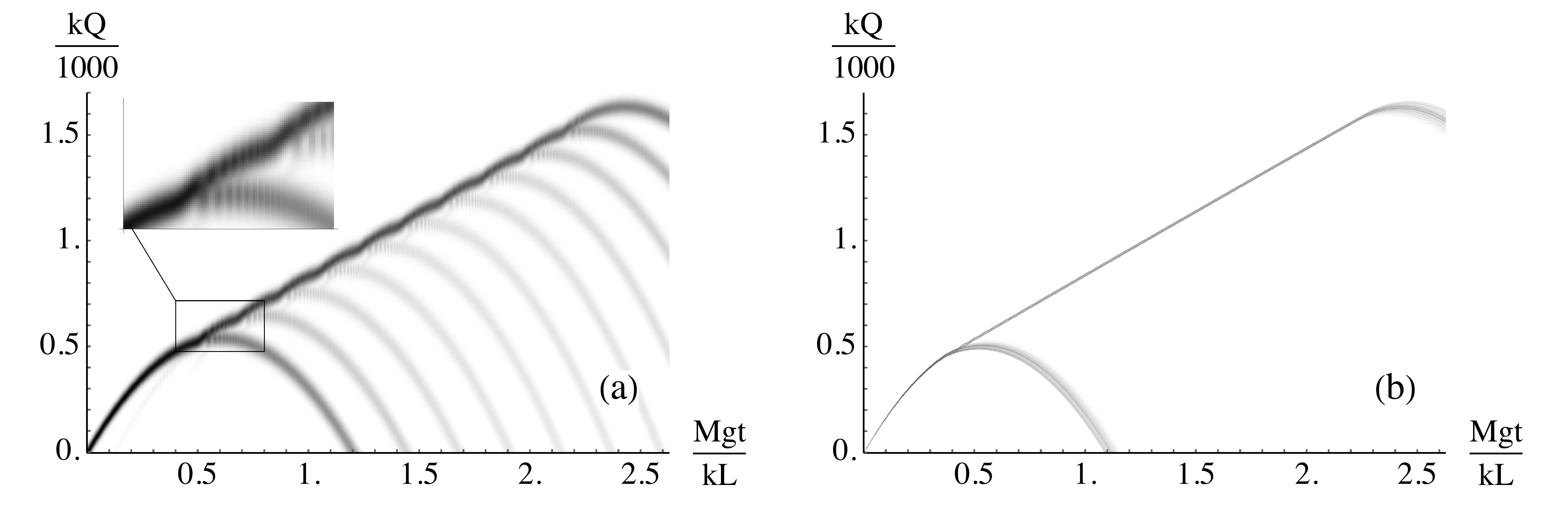}
\caption{\label{F:positionSpaceDist}
The probability distribution in position for the quantum system and a classical ensemble. Both figures have the same axes and parameters $\alpha$, $\beta$ and $\epsilon$, and the same initial $L_{z}/L=\sqrt{5/6}$ as the trajectories shown in the classical figure Fig.~\ref{fig:ClassEvol1}. The initial quantum wave packet has the same average values of $Q$ and $P$ as the classical initial values; see the text for discussion of the effects of quantum uncertainties in $Q$ and $P$. The color scheme is normalized to show highest probability amplitude as black and zero probability amplitude as white.\\
(a): The quantum probability distribution of the weight's position $\pr(Q,t)$ as a function of time, incoherently summed over $\ket{m}_{f}$. $L/\hbar = \sqrt{30}=\sqrt{l(l+1)}$ for $l=5$ in the quantum case shown here, whereas it is infinite in all classical cases. We see a single wave packet splitting coherently into a multi-branched superposition of trajectories with longer and shorter durations of downconversion. The \textbf{inset} shows the first two crossings enlarged to exhibit the oscillations of probability between different branches during crossings.\\
(b): The classical ensemble density in position. An ensemble of 1000 trajectories is chosen, where the initial phase $\phi(0)$ is uniformly distributed in the interval $[0,2\pi)$ and all other initial conditions are chosen as in Fig.~\ref{fig:ClassEvol1}. All trajectories are then numerically evolved in time and we count the position in 1700 equally sized bins at each of 1312 time steps. The gray scale is not linear but shows the square root of the binned probability distribution to make features more clearly visible.}
\end{figure*}

In Fig.~\ref{F:positionSpaceDist} we depict an example of the quantum weight's time evolution. This was obtained by numerically solving \eqref{E:ReducedSchroedinger}, inserting the resulting $\psi_{m}(P,t)$ into (\ref{ReductionMapping}) to obtain $\Psi_{m}(P,t)$, and then Fourier transforming to obtain the probability distribution of the weight's position $Q$ summed over $m$ states,
\begin{equation}
	\pr(Q,t)= \sum_{m=-l}^l|\langle m|\langle Q|\Psi(t)\rangle|^2\;,
\end{equation}
as a function of time. The initial state was a wave packet with $m=l$ as in \eqref{packet}, with position width $20/k$ (momentum width $D=\hbar k/20$) and initial mean momentum $3M\Omega/k = 0.6 kL$.  Comparing Fig.~\ref{F:positionSpaceDist}  to Fig.~1, we see that the quantum motion of the daemon-driven weight is a remarkably simple probabilistic mixture of \textit{both} the downconversion and decoupling classical phases.

Some of the quantum daemon's further features become more apparent in the weight's momentum distribution $\pr(P,t)$, obtained from \eqref{eq:ansatz1} directly, and shown in Figure \ref{F:momentumSpaceDist}. Initially launched upwards, the rising weight decelerates under gravity until it has slowed to the \textit{quantum} critical velocity $v_{q}=\Omega/k-\hbar k/(2M)=v_{c}-\hbar k/(2M)$. At this velocity, the energy $\hbar\Omega$ of the $\hat{L}_{-}$ transition matches the kinetic energy change if the weight momentum should increase by $\hbar k$.  Since such a momentum jump is precisely the operation of $e^{i k\hat{Q}}$, the nonlinear coupling between subsystems which are otherwise badly mismatched in frequency becomes a resonant coupling just at this quantum critical speed, and the result is that the probability distribution forks. With some probability, the weight simply continues to decelerate under gravity, following the familiar parabolic trajectory. With greater probability, however, the weight makes a quantum jump of $\hbar k$ in upward momentum. Since this corresponds to a velocity jump to $v_{q} +\hbar k/M = v_{c} + \hbar k/(2M)$, the momentum jump is visible as a kink in the curve of the position wave packet's motion in Figure~\ref{F:positionSpaceDist}. 

\subsubsection{Periodic momentum kicks}
After each quantum jump in the quantum weight's momentum, gravitational deceleration at the rate $\dot{P}=-Mg$ continues. After the time interval $\Delta t = \hbar k/(Mg)$, therefore the weight has slowed down from $v_{q}+\hbar k/M$ to $v_{q}$ again---and the forking repeats. The probability of the momentum jumping up from $Mv_{q}$ to $Mv_{q}+\hbar k$ is even higher in forkings after the first one, as we will work out in the following section (see eqn.~\eqref{prm}), and so there is a substantial probability that the weight will continue to rise against the linear force, at the \textit{same} time-averaged velocity $[v_{q}+(v_{q}+\hbar k/M)]/2 =v_{{c}}=\Omega/k$ as in the classical case, until at most $l$ jumps have occurred (since $\hat{L}_{-}$ simply annihilates $|-l\rangle$). After this point, with whatever probability remains, the weight accelerates downwards. Since $l$ may be made arbitrarily large, however, the weight can in principle be lifted to arbitrary height. Examining the full quantum state $\ket{\Psi(t)}$, we can confirm that the work done on the weight is exactly matched by the energy lost from the high-frequency subsystem, and that each jump in the weight's velocity occurs with an angular momentum transition from $|m\rangle_{f}$ to $|m-1\rangle_{f}$. The daemon is like an engine that consumes its fuel quantum by quantum.

\begin{figure*}
\centering
\includegraphics[width=\textwidth]{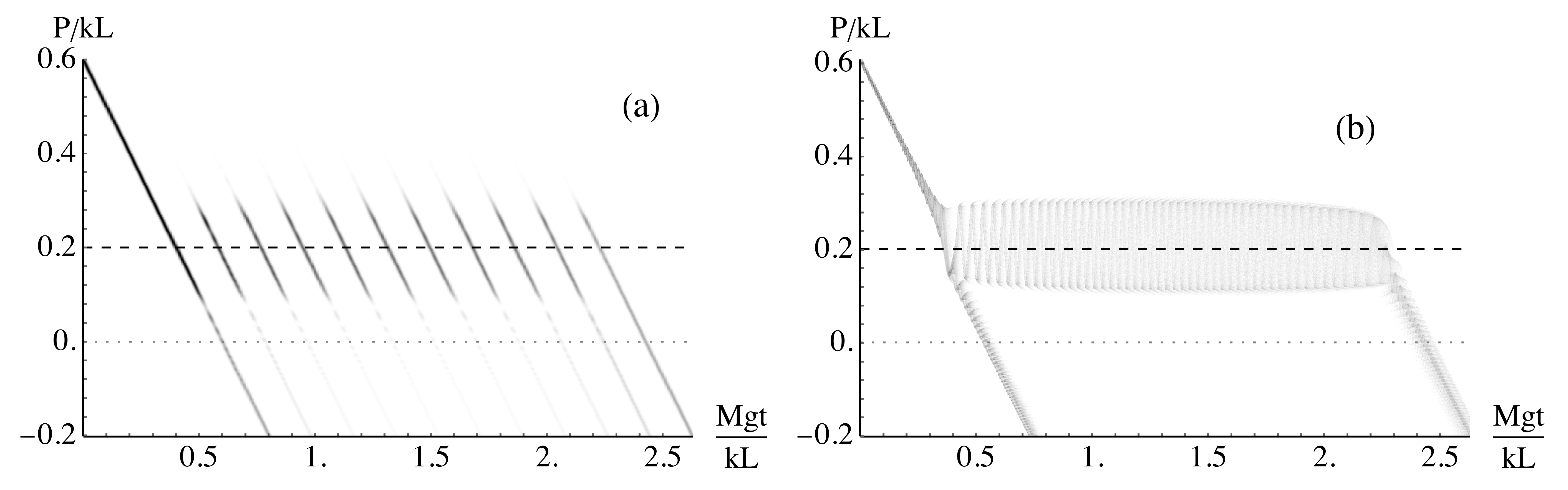}
\caption{\label{F:momentumSpaceDist}The probability distribution of the weight's momentum for the quantum system and a classical ensemble. Both figures have the same axes and parameters $\alpha$, $\beta$ and $\epsilon$, and the same initial $L_{z}/L=\sqrt{5/6}$ as the trajectories shown in the classical figure Fig.~\ref{fig:ClassEvol2}. The color scheme is normalized to show highest probability density as black and zero probability density as white. In contrast to Fig.~\ref{F:positionSpaceDist}.\\
(a): The quantum probability distribution of the weight's momentum $\pr(P,t)$ as a function of time, incoherently summed over $\ket{m}_{f}$, for the same evolution as shown in Figure~\ref{F:positionSpaceDist} (a). The oscillations of $P(t)$ around $0.2$ with period $2\pi/\omega\doteq 0.034 kL/(Mg)$ that appeared in Fig.~\ref{fig:ClassEvol2} are replaced entirely by periodic quantum jumps at time intervals of $\Delta t=\hbar k /(Mg)\doteq 0.18 kL/(Mg)$. The gray scale here is linearly proportional to the probability density itself, and not its square root; this shows more accurately that the intermediate quantum branches are of quite low probability.\\
(b): The classical ensemble distribution of the weight's momentum. The ensemble chosen is the same as in Fig.~\ref{F:positionSpaceDist}. The position is counted in 1700 bins over 1968 time steps. The result shows oscillations in the downconversion phase that are a somewhat smeared version of those seen in Fig.~\ref{fig:ClassEvol2}. The gray scale shows the square root of the binned probability distribution as in Fig.~\ref{F:positionSpaceDist}.}
\end{figure*}

The regular succession of wave function forkings is the periodic operation of the quantum daemon, surprisingly analogous to the cyclic operation of a macroscopic engine. It is likewise analogous to the nearly periodic motion within potential basins in the classical tilted washboard representation, as described in \cite{Us_PRE}, and to the orbits inside the separatrix on the $\mathbf{L}$-sphere as discussed above in Section III. Note, however, that the time between successive quantum jumps $\Delta t=\hbar k/(Mg)$ has nothing to do with the classical period of bound oscillations in the downconversion phase $\omega^{-1}\sim \sqrt{M/(\gamma L k^{2})}$. This time scale difference can be seen clearly in Fig.~6

From $\pr(P,t)$ in Figure \ref{F:momentumSpaceDist} we can also see that the daemon induces true jumps in the weight momentum, and not just short bursts of high acceleration. The instant at which the velocity jump occurs is probabilistically spread over a short continuous interval, but there is never any probability, during this interval, for the weight velocity to take any intermediate values. It is expected that a quantum system that is slowly driven, as our fast subsystem is by our weight, will exhibit such energy jumps; see, for example, Figure~8 of Ref.~\cite{cohen_2006}. Because our entire system is closed, we see here the corresponding back-action jumps of the slow weight. 

(The fact that the daemon's momentum jumps of $\hbar k$ are as large as its average momentum $Mv_{c}=Mk/\Omega$ is \textit{not} generic, but has been chosen simply to provide clear plots with a momentum axis that starts at zero. The high frequency $\Omega$ has dropped out of $\hat{H}_{\mathrm{eff}}$, and so raising $\Omega$ to be larger than $\hbar k^{2}/M$ would effectively do nothing but shift Fig.~\ref{F:momentumSpaceDist} upward, raising the average momentum $Mv_{c}=M\Omega/k$ arbitrarily while keeping the momentum jump unchanged at $\hbar k$.)

\subsubsection{Compensating for dispersion}
Besides steadily lifting the weight, the quantum daemon's momentum kicks also have another remarkable effect. Because faster parts of the weight's wavepacket are later to slow down to the critical velocity, while trailing parts of the wave packet fall to the critical velocity sooner, the trailing parts of the wave packet get kicked up to higher speed before the faster parts receive their kicks. These earlier kicks make the trailing parts of the wave packet catch up again, and so each kick effectively undoes the gradual dispersion that the wavepacket otherwise experiences in the linear external potential in position space. As a result the width of the position wavepacket in the downconversion branch (\textit{i.e.} the rippled spine along the top of the density pattern shown in Fig.~\ref{F:positionSpaceDist}) stays almost constant throughout quantum daemon operation. Only the falling branches below it, corresponding to failed adiabatic transitions, show the expected packet broadening during free fall.

\subsubsection{Quantum stalling}
Along with the possibility of raising the weight, the multiple probability branches of Figs.~\ref{F:positionSpaceDist} and \ref{F:momentumSpaceDist} show that each velocity jump may also fail to occur. We can understand the forking of the weight's probability distribution analytically in the extreme quantum limit, where $l\gamma \ll\hbar k^{2}/M$ as in Fig.~\ref{F:positionSpaceDist}, by applying Landau-Zener post-adiabatic theory \cite{Zener, Landau}. This will show that the possibility of the weight's wave packet taking a lower branch at some point cannot be entirely eliminated for any parameter choice, implying a limitation on daemon efficiency similar in implication to the one we found in the classical case \cite{Us_PRE}, yet very different in origin.

\subsection{Comparison with a classical ensemble}
It is arguably inappropriate to compare quantum daemon evolution with any single classical trajectory, because individual classical trajectories all have definite values of their initial phase space variables. Our initial quantum wave packet is quite localized in $Q$ and $P$, and it is an exact eigenstate of $\hat{L}_{z}$; but as an eigenstate of $\hat{L}_{z}$ it can be said to consist of a superposition of all values of $\phi$.

We therefore also consider an ensemble of $N=1000$ classical systems. All the trajectories have the same initial values of $P$ and $Q$, which are equal to the initial expectation values of $\hat{Q}$ and $\hat{P}$ in the quantum evolution that we have just described, as well as to the corresponding initial values in the two classical trajectories shown in Figs.~1, 2 and 4. All the ensemble trajectories also have the same initial value of $L_{z}=\sqrt{5/6}L$ as in the quantum case and the previous classical trajectories. In this classical comparison ensemble, however, the initial values of the phase $\phi$ are distributed evenly over the interval $[0,2\pi)$. This classical ensemble thus corresponds more closely than any individual classical trajectory to our chosen initial quantum state.

We then evolved this ensemble under our system's classical equations of motion. To obtain probability densities which can be compared to the quantum probability densities, we defined uniformly binned intervals  in $Q$ and $P$, and examined the ensemble at a series of equally spaced times. At each of these times we counted the number of trajectories within each bin, and convert this whole number into a gray scale level. This yielded an array of grayscale pixels in either $Q,t$ or $P,t$. 

The result are shown in the (b) panels of Figure~\ref{F:positionSpaceDist} and Figure~\ref{F:momentumSpaceDist}, beside the corresponding quantum probability densities in the (a) panels. We can note immediately that these classical ensemble plots present essentially the same features as the individual trajectory plots in  Figure~\ref{fig:ClassEvol1} and Figure~\ref{fig:ClassEvol2}. Since the exact evolution of the classical system depends on the initial value of the phase $\phi$, however, the classical ensemble soon becomes noticeably `smeared' in both $Q$ and $P$.

The noticeably greater width of the quantum probability distributions, in contrast, reflect the initial widths of the quantum wave packet. The comparison quantum ensemble is initially distributed over all $\phi$ but has unique initial values of $Q$ and $P$. One might therefore ask how a classical ensemble would evolve if it also began with a finite-width Gaussian distribution of $Q$ and $P$ values. 

In fact this question is easy to answer. The classical evolution has time translation symmetry, such that for any solution $\mathbf{R}(t)=(Q(t),P(t),\phi(t),L_{z}(t))$ and any constant $t_{0}$, $\mathbf{R}(t-t_{0})$ is also a solution. At early times long before the system approaches the nonlinear resonance, however, the evolution is simply that the weight decelerates under gravity, $L_{z}$ remains constant, and $\phi$ moves around the circle $[0,2\pi)$ at a uniform rate. Insofar as our $N$-element ensemble (with $N=1000$) approximates a uniform distribution over all initial $\phi$ values, it is invariant under the steady translation of $\phi$, and hence shifting the initial momentum $P\to P+\delta P$ at any initial time that is long before resonance corresponds simply to a shift in time $t\to t-\delta P/(Mg)$, plus a compensating shift in $Q$.

The classical evolution also has an obvious translation symmetry, such that for any solution $Q(t),\phi(t)$ and any constant $a$, the shifted evolution $Q(t)+a,\phi(t)+a$ is also a solution. A uniform ensemble of initial $\phi$ values is again invariant under such a shift, and so shifting or smearing the initial value of $Q$ in the ensemble simply shifts the entire ensemble evolution in $Q$, rigidly. 

Hence the additional effects of smearing the classical initial conditions on $Q$ and $P$ will simply be to smear Figs.~5b) and 6b) horizontally, and to further smear Fig.~5b) vertically. One can therefore accurately picture the evolution of a classical ensemble whose initial phase space distribution exactly matched the quantum one, simply by blurring Figs.~5b) and 6b) to make the initial distributions match the quantum ones. Insofar as the greater breadth of the quantum distributions is a quantum effect, therefore, it is a very simple quantum effect that is essentially due to the quantum initial conditions, and not to qualitative differences between quantum and classical time evolution. The evolution of the quantum daemon after it encounters the resonance, however, exhibits dramatically quantum mechanical effects.

\subsection{Post-Adiabatic Analysis}\label{sec:analytics}

We can understand the quantum evolution generated by the $\hat{H}_\mathrm{eff}$ of (\ref{E:qEffectiveH}) by examining its instantaneous eigenspectrum. In dimensionless terms this means examining the matrix elements $h_{mn}(t)$ defined in \eqref{E:ReducedSchroedinger}. The time-dependent part $h_{m}(t)$ is diagonal in the $|m\rangle$ basis; the non-diagonal part ${w}_{mn}$ is time-independent, and is non-zero only for $m-n=\pm1$. In the limit where $\gamma l\ll\hbar k^{2}/M$, all the elements of ${w}_{mn}$ are small, while the differences between neighboring eigenvalues of $h_{m}(t)$ are mostly of order unity. The adiabatic approximation and perturbation theory are therefore in most cases excellent. Together they imply that the system will have negligible amplitude to make any transitions between different $|m\rangle_{f}$ states; this amounts to adiabatic decoupling, and failure of the quantum daemon to perform steady downconversion.
\begin{figure}[htb]
\centering
\includegraphics[width=0.475\textwidth]{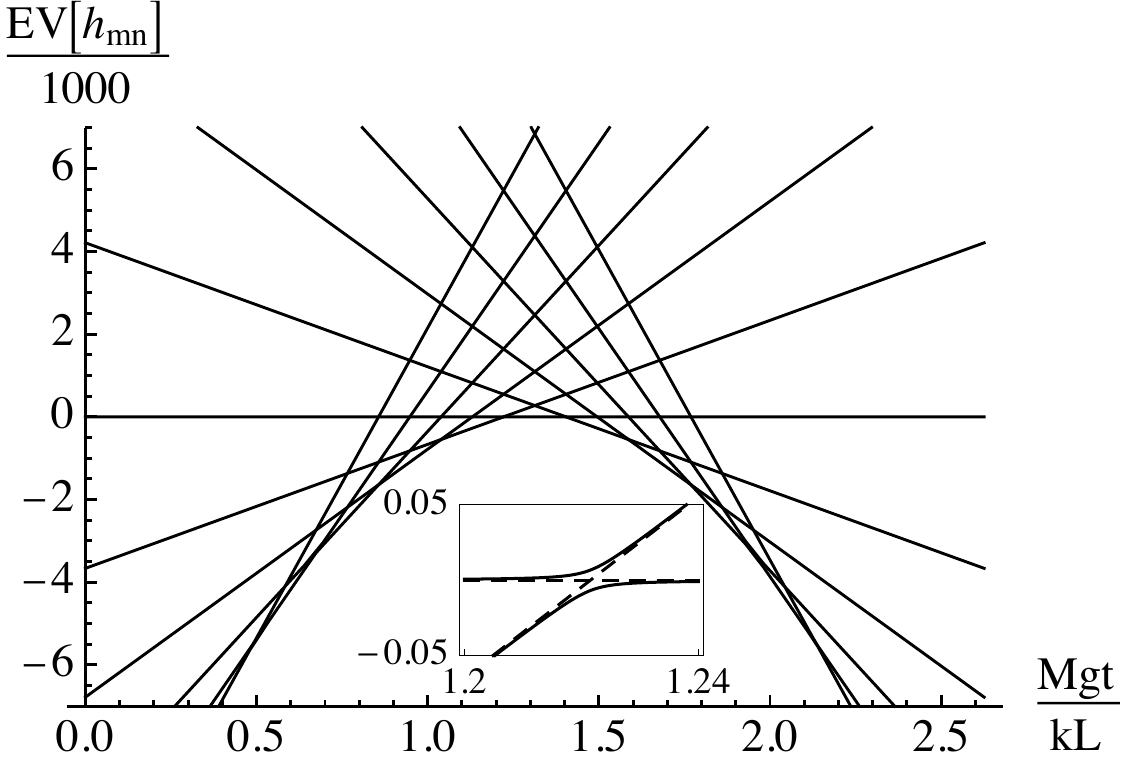}
\caption{Instantaneous eigenvalues of $h_{mn}(t)$ from \eqref{E:qEffectiveH}, for parameters as in Figs.~\ref{F:positionSpaceDist} and \ref{F:momentumSpaceDist}. \textbf{Inset:} zoom to show an avoided crossing. Dashed lines are $h_{0}(t)$ and $h_{1}(t)$ from \eqref{E:qEffectiveH}.}
\label{fig:energylevels}
\end{figure}
\subsubsection{Avoided crossings}
The loophole in that conclusion, however, is another subtle issue of resonance, this time quantum mechanical. In fact it is not quite always true that the eigenvalues of $h_{m}(t)$ are separated by gaps of order unity. For certain values of $t$, some $h_{m}$ eigenvalues actually cross. In particular, whenever $2Mg(t_{P}-t)/(\hbar k)$ is an integer $n$, levels $m$ with equal $|m-n/2|$ are degenerate. If $\gamma l M/(\hbar k^{2})$ is small, the eigenvalues of $h_{m}$ are close to those of $h_{mn}$. As long as $\gamma\not=0$, however, the eigenvalues of $h_{mn}$ never actually cross---the crossings of $h_{m}$ eigenvalues are \textit{avoided} by the eigenvalues of $h_{mn}$, as ${w}_{mn}$ lifts the degeneracy in $h_{m}$. See Fig.~\ref{fig:energylevels}. 

If $\gamma l M/(\hbar k^{2}) = \sqrt{l/(l+1)}\tilde{\gamma}\tilde{M}(L/\hbar)^{2}$ is small, then the intervals of $t$ within which ${w}_{mn}$ has this small but potentially significant effect on the eigenspectrum of $h_{mn}$ are all brief compared to the much longer time spans between $h_{m}$ crossings. We can therefore conclude that in the extreme quantum limit $\gamma l M/(\hbar k^{2})\ll 1$ the quantum time evolution is entirely adiabatic, except possibly in the brief episodes around times $t=t_{P}-n \hbar k/(2Mg)$. During these brief potentially non-adiabatic episodes, the only non-adiabatic evolution which can actually occur will consist of transitions between the nearly degenerate pairs of $|m\rangle_{f}$ eigenstates that are crossing.

Moreover, the non-diagonal Hamiltonian part ${w}_{mn}$ directly couples only $|m\rangle_{f}$ of which the $m$ values differ by one. Only some of the $h_{m}$ level crossings are between such states (the lowest `arc' in Fig.~\ref{fig:energylevels}). The degeneracies between other states will also be lifted by the ${w}_{mn}$ perturbation, but only at higher orders in perturbation theory. If first order in $\gamma l M/(\hbar k^{2})$ is already small, then these higher order level splittings will be much smaller still. 

Let us therefore first consider only the potentially non-adiabatic evolution around avoided crossings between states $|m\rangle_{f}$ and $|m-1\rangle_{f}$, within the time interval $t=t_{p}-(2m-1)\hbar k/(2Mg)+\delta t$ for $|\delta t|\ll\hbar k/(Mg)$. The projection of $\hat{H}$ into this nearly degenerate subspace is then
\begin{align}\label{Hproj}
\hat{H}_{\mathrm{pr}}=& \frac{\hbar^{2}k^{2}}{2M}\left[m(1-m)+\frac{Mg}{\hbar k}\delta t\right]\hat{I}\\
&+\frac{\hbar k g}{2}\delta t\,\hat{\sigma}_{z}-\frac{\hbar \gamma}{2}\sqrt{l(l+1)-m(m-1)}\hat{\sigma}_{x}\;.\nonumber 
\end{align}
where $\hat{I}$ is the identity operator in the two-state subspace, and the Pauli operators are defined for each $m$, $m-1$ pair as
\begin{align}
\hat{\sigma}_{z}=&|m\rangle\langle m| - |m-1\rangle\langle m-1|\nonumber\\
\hat{\sigma}_{x}=&|m\rangle\langle m-1| + |m-1\rangle\langle m|\;.
\end{align}
\subsubsection{The Landau-Zener problem}
The time-dependent Hamiltonian \eqref{Hproj} in the two-dimensional Hilbert space is the classic problem solved independently by Landau and Zener in 1932 \cite{Landau, Zener}. The instantaneous eigenstates of $\hat{H}_{\mathrm{pr}}$ change continuously in time, such that the lower energy state is $|m\rangle$ for $\delta t\ll - \gamma \sqrt{l(l+1)-m(m-1)}/(kg)$, but $|m-1\rangle$ for $\delta t\gg + \gamma\sqrt{l(l+1)-m(m-1)}/(kg)$. The higher energy state changes oppositely. Under the adiabatic approximation, therefore, the evolution through the avoided crossing will be for an initial $|m\rangle$ to evolve into $|m-1\rangle$, while $|m-1\rangle$ evolves into $|m\rangle$.

Computing the exact time evolution, however, Landau and Zener showed that the probability that the system will actually evolve oppositely to the adiabatic approximation, such that initial $|m\rangle$ or $|m-1\rangle$ states emerge unchanged except by a phase, is exactly
\begin{equation}\label{prm}
\pr_{m}=\exp\left(-\pi\frac{\gamma^{2}[l(l+1)-m(m-1)]}{2 kg}\right).
\end{equation}
The probability of the adiabatic evolution, in which $|m\rangle$ and $|m-1\rangle$ switch, is then $1-\pr_{m}$. In the adiabatic limit $kg/\gamma^{2}\to0$ the non-adiabatic probability becomes non-analytically small, but for finite $kg/\gamma^{2}$ it is never zero.

Turning now to the other avoided crossings, above the  lowest arc in Fig.~\ref{fig:energylevels}, we can in principle also perform a Landau-Zener computation to determine the exact evolution through them. The role of the $\gamma \hat{\sigma}_{x}$ term in their cases, however, will be played by an $n^{\mathrm{th}}$ order effective Hamiltonian term obtained by pursuing perturbation theory in $\gamma$ far enough to obtain a coupling between the $|m\rangle$ and $|m-n\rangle$. Unless $kg$ is extremely small indeed, the results will give $\pr_{m}$ extremely close to 1 for these higher crossings. There is therefore a wide regime in which the higher crossings are effectively not avoided, and time evolution proceeds diabatically through them following the straight lines in Fig.~\ref{fig:energylevels}. This is effectively the case in the evolution shown in Figs.~\ref{F:positionSpaceDist} and \ref{F:momentumSpaceDist}. Re-plotting these figures with a highly nonlinear gray scale reveals additional branches of very low probability, produced by adiabatic transitions through the higher-order avoided crossings.

\subsubsection{Landau-Zener and the quantum daemon}
In Figs.~\ref{F:positionSpaceDist} and \ref{F:momentumSpaceDist} we start with the fast sector in the state $|m\rangle=|l\rangle$, which is the initial ground state of the eigenspectrum of (\ref{E:ReducedSchroedinger}). The system remains in this $|l\rangle_{f}$ state adiabatically, while the weight's wavepacket accelerates downward, until the first avoided crossing in Fig.~\ref{fig:energylevels} is encountered; the different $P$ components of the wavepacket encounter this crossing at displaced times, but the momentum width of the packet is small enough that this spreading is hardly noticeable. What happens at this first avoided crossing is a bifurcation of the wave packet due to non-adiabatic Landau-Zener evolution through the avoided crossing. The bifurcation is into two branches---diabatic and adiabatic.

With probability $\pr_{l}=\exp[-\pi l^{2}\gamma^{2}/(kg)]\doteq 0.31$ the Landau-Zener evolution through this first avoided crossing is diabatic, and then since all the higher crossings are also diabatic for these parameters, the $|l\rangle_{f}$ branch of the system will, at least to first order, not produce further entanglement with other states $|l'\rangle_{f}$; no downconversion occurs, the total quantum state within this superposed branch remains a tensor product of fast and slow sector states (\textit{i.e.} it does not branch further), and the corresponding branch of the weight's wave packet continues to accelerate downwards. This is the first of the parabolic branches in Fig.~\ref{F:positionSpaceDist} and the first continuing downward branch in Fig.~\ref{F:momentumSpaceDist}. Note that we have deliberately chosen our parameters to make $\pr_{l}$ large enough for the diabatic branches to be seen in these plots, but all the $\pr_{m}$ can in principle be arbitrarily small. 

With probability $1-\pr_{l}\doteq 0.69$, however, the Landau-Zener evolution through the first avoided crossing is adiabatic, dropping from $|l\rangle_{f}$ to $|l-1\rangle_{f}$. The weight's wavepacket receives an upward kick $\hbar k$ in momentum, jumping discontinuously upward in Fig.~\ref{F:momentumSpaceDist}. Shortly after this first avoided crossing, therefore, the total quantum state of the system is a superposition of two branches, which differ in both $l$ and weight momentum. Although we have quoted Landau-Zener probabilities, in fact the total evolution remains unitary and the bifurcation is coherent. The fast and slow sectors of the total system are thus quantum mechanically entangled. 

\begin{figure}[htb]
\centering
\includegraphics[width=0.475\textwidth]{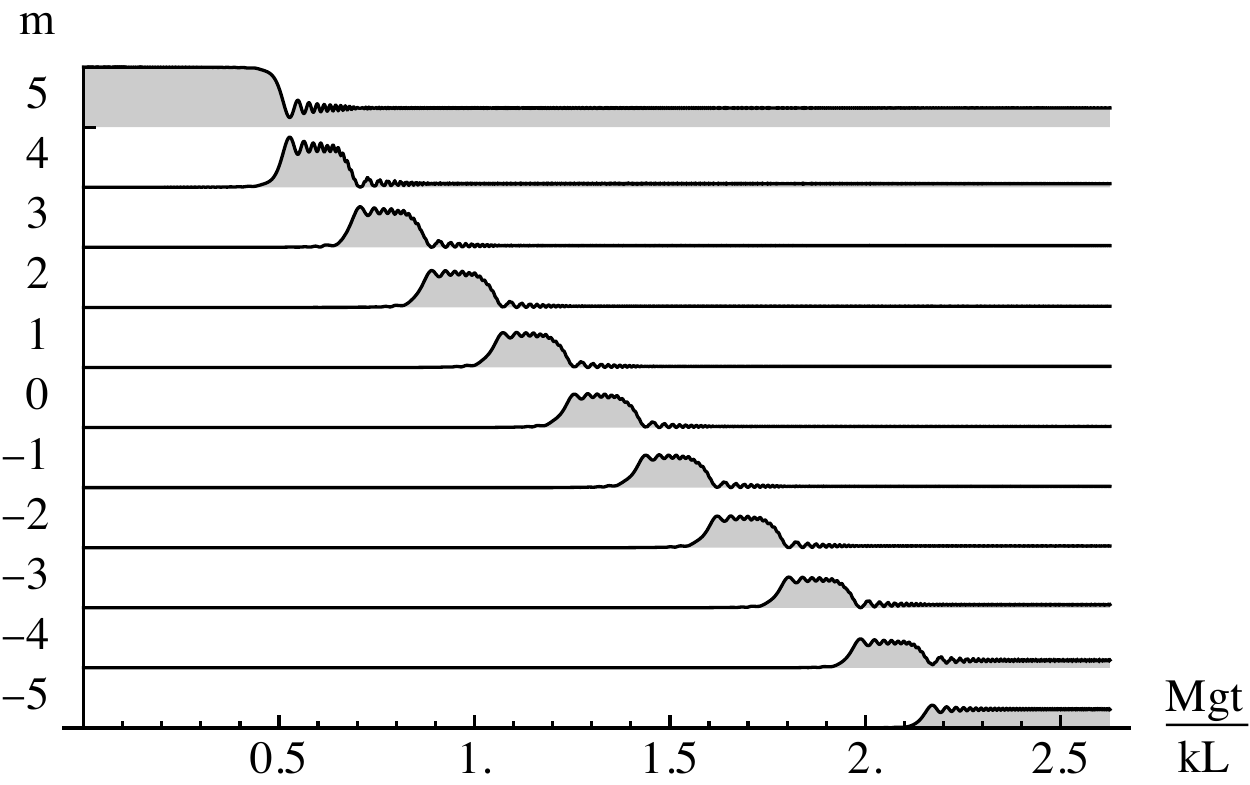}
\caption{\label{F:occupationProbability}The reduced probabilities $\langle\psi_{m}(P_{0},t)|\psi_{m}(P_{0},t)\rangle$ as functions of time. The numbers on the vertical axis indicate the $m$ value to which the corresponding horizontal strip refers; within each horizontal strip the vertical axis is probability from 0 to 1, and the area under each curve is shaded as a guide to the eye.}
\label{fig:strips}
\end{figure}

It is worth noting, on the other hand, that although the asymptotic consequence of the crossing episode are simply the probabilities $\pr_{m}$ and $1-\pr_{m}$ of the two crossing levels, the non-adiabatic evolution that produces them is non-trivial, and is not really instantaneous. The probability amplitude actually oscillates back and forth between the two levels several times. This can be seen well in Fig.~\ref{F:occupationProbability} which shows the probability for the fast subsystem to occupy a particular $\ket m$ state over time. There we see that for the parameters of Figs.~\ref{F:positionSpaceDist} and \ref{F:momentumSpaceDist} the amplitudes never actually settle down completely to their asymptotic values before the next avoided crossing is encountered. These oscillations can also be seen clearly in the inset of Fig~\ref{F:positionSpaceDist}. The simple picture of smooth adiabatic evolution punctuated by instantaneous Landau-Zener transitions is thus only an approximation for the case we have shown. For smaller $g$, the evolution would be more adiabatic and the simple picture would be accurately realized. Because the Landau-Zener formula is the exact non-perturbative prefactor of a convergent post-adiabatic perturbation series, however, we can see that the approximation of isolated Landau-Zener transitions gives the branching ratios surprisingly well, even in the case we have shown.

\subsubsection{Multiple Landau-Zener transitions}
After the first avoided crossing, adiabatic evolution continues in both branches of the daemon evolution, with $|m\rangle_{f}$ conserved and the weight decelerating. The weight's upward speed is still positive initially---and on the adiabatic branch it has even been kicked higher---so the weight continues to rise for a while. In the diabatic branch, the negative acceleration will soon bring the weight's upward velocity below zero and it will then fall forever. In the adiabatic branch, however, the upward-kicked weight is still rising when the reduced system, proceeding along the lowest arc of energy levels in  Fig.~\ref{fig:energylevels}, encounters another avoided crossing through which the probability of adiabatic evolution is not negligibly small. The bifurcation that occurred at the first avoided crossing is then repeated, with the slightly different branching probability $\pr_{l-1}$. This continues until the last avoided crossing leaves the adiabatic branch in $|-l\rangle$. After this final avoided crossing has been passed, the further evolution is adiabatic for all branches of what is finally a multi-branched Schr\"odinger's Cat state with $2l+1=11$ branches. This final state has high quantum entanglement between the fast and slow sectors of the total system.

We repeat here the observation that we made in Section II.C above, that quantum and classical systems can be adiabatic in different ways. In the classical evolutions shown in Figs.~1, 2, and 5b) and 6b), the transitions between decoupling and downconversion phases were necessarily non-adiabatic. We have just seen here, however, that the quantum daemon beginning in an analogous initial state can proceed from decoupling through complete downconversion and return to decoupling, all while making only adiabatic Landau-Zener transitions. This represents another subtle but important difference between the quantum and classical dynamics of Hamiltonian daemons, at least in the strong quantum regime.

Generalizing our quantum results beyond the strong quantum regime of $l\gamma \ll\hbar k^{2}/M$ becomes complicated quickly, because the intervals of non-adiabatic evolution around the crossings can extend to encompass multiple crossings, and because the higher-order crossings also become important, leading to a complex flow of interfering quantum amplitudes through the entire web of $h_{n}$ levels. Detailed investigation of this complex regime must be left for future study. For now we can report only a qualitative impression from preliminary numerical analysis in a range of regimes: the quantum daemon appears to operate quite generally and robustly, but always with at least some small degree of inefficiency and uncertainty. 

\section{Discussion}\label{sec:discussion}
The undulating uppermost branch shown in Figure \ref{F:positionSpaceDist}a) is a quantum analog to the constant mean velocity in the downconversion phase of the classical Hamiltonian daemon as shown in \cite{Us_PRE} and reviewed in Section II above. Comparing Figs.~5a) and 6a) with 5b) and 6b)---or with Figs.~1 and 2---reveals strong similarities between the quantum and classical systems. The times at which downconversion can first begin and must eventually end are the same in both classical and quantum cases, and so is the steady speed with which the weight rises against gravity during downconversion. These similarities reflect the fact that three out of four dimensionless parameter ratios (\textit{i.e.}, all except $L/\hbar$) are the same in both cases. The strong resemblance between the quantum and classical daemon would seem to confirm that Hamiltonian daemons are a generally Hamiltonian phenomenon with both quantum and classical limits.

The differences between Figs.~\ref{fig:ClassEvol1} and \ref{fig:ClassEvol2} and Figs.~\ref{F:positionSpaceDist}a) and \ref{F:momentumSpaceDist}a) are just as striking as the similarities, however. Instead of two types of trajectories, which either contain the downconversion phase or do not contain it, the quantum system features a many-branched quantum superposition of both phases. In this respect it resembles the classical ensembles shown in Figs.~5b) and 6b), even though it starts in a single pure quantum state.

Furthermore, instead of the continuously decreasing high-frequency energy of the classical daemon, the quantum daemon consumes its fuel quantum by quantum. And instead of the quasi-periodic oscillations of the classical downconversion phase, the quantum daemon lifts the weight through a series of quantum jumps in momentum. These quantum features go beyond the ensemble-like broadening of initial conditions that is required by the Uncertainty Principle, and represent radically non-classical forms of dynamical evolution. We have thus shown that the basic phenomenon of daemon dynamics---steady downconversion within a small, closed system---appears in quantum mechanics quite generally, even well beyond the semi-classical regime. The detailed mechanism by which quantum daemons work can be quite different from the classical mechanism, however, and the behavior of quantum Hamiltonian daemons can be thoroughly quantum.

\subsection{Efficiency limits}
In particular, when we consider the intrinsic efficiency limit on the extreme quantum daemon we discover that it is quite different conceptually from the limitations on classical daemons, which were based on phase space volumes. The classical limitation on the daemon was simply the fact that, due to adiabatically conserved orbit area, the system would in general be expelled from its downconversion bound orbit before it could transfer all the fast energy to the weight. Until this time of expulsion, the classical system's downconversion phase persisted stably. In terms of dynamics on the sphere this corresponded to the final $L_{z}$ at the end of a  downconversion phase being approximately equal to \textit{minus} the initial $L_{z}$ value before the downconversion phase \cite{Us_PRE}.

In the quantum case, in contrast, there is no limitation strictly forbidding the system to convert all energy from fast to slow. In fact there is always a chance, depending on parameters, for this to happen. There is also a finite probability, however, for the system to cease downconversion and revert to the decoupling phase, through a diabatic Landau-Zener transition at any avoided crossing. The quantum daemon enjoys its possibility of 100\% efficiency, in a lucky run, at the cost of having a finite chance that downconversion will cease well before the time at which a classical daemon would stall. The average efficiency of quantum daemons can only attain unity in the perfectly adiabatic limit $Mg\to0$, at which point the daemon is producing zero power. We emphasize, however, that quantum daemons are not necessarily bad daemons: the average efficiency of the quantum daemon can in principle be arbitrarily close to unity, for sufficiently large $\gamma^{2}/(kg)$. 

The classical and extreme quantum limitations also differ in another curious way. Classically, the probability to enter the downconversion phase becomes smaller, the further $L_{z}(0)$ lies below $L$; the chance of downconversion is one for $L_{z}(0)$ above a certain threshold close to $L$, and it drops smoothly towards zero for lower $L_{z}(0)$ \cite{ignition_eprint}. For the quantum daemon in the extreme quantum limit, the initial state with maximum $L_{z}$ is favored even more heavily than this, though. If the system begins with an $m<l$, it proceeds adiabatically through Fig.~7 from left to right along an energy line higher than the lowest arc. When it meets the lowest arc in an avoided crossing, the chance that the system will jump diabatically onto the lowest-arc state, and thus begin downconversion, is the same as the chance that the system would at that point have jumped diabatically \textit{out} of the lowest-arc state, if it had approached that avoided crossing on the lower branch. Hence if the quantum daemon is to operate efficiently from an initially ``full fuel tank'' $m=l$, with low chance of spontaneously stalling before all fuel is consumed, then the chance that the daemon will perform any downconversion at all, if it starts with $m<l$, must be correspondingly low, too. Highly efficient quantum daemons in the extreme quantum limit are in this sense highly ``fussy'': they work well if they start with full fuel, but hardly at all if their initial fuel level is even one quantum short of full. A less fussy extreme quantum daemon, which is willing to work with a wider range of initial $m$, is inevitably also more likely to stop working, spontaneously, even when it has plenty of fuel left.

\subsection{A microscopic precursor to thermodynamical entropy?}
In \cite{Us_PRE} we noted that phase space volumes which cannot shrink are associated with entropy in statistical mechanics, and speculated that the somewhat different kinds of growing, shrinking, or conserved phase space volumes which explain the efficiency limits of classical daemons could represent previously unsuspected microscopic precursors to thermodynamic entropy. In the quantum case that we have now studied here, we can also relate the fundamental efficiency limits of Hamiltonian daemons to a process that is often invoked to explain macroscopic thermal inefficiency, namely effective loss of information as it is passed from slow, observed degrees of freedom to other degrees of freedom that are unobservably fast. 

Although we have described probability distributions for the lifted weight, the evolution of our full system is quantum mechanically unitary, and the final state is really a coherent `Schr\"odinger's Cat' superposition of all different fast sector energy levels, with the corresponding wave packets for the weight all at correspondingly different heights. If the system starts in a direct product of fast and slow sectors, therefore, quantum entanglement between the sectors grows steadily with every quantum of energy that is transferred---or coherently not transferred!---by the daemon from the fast to the slow subsystem. 

Observing the coherence of this Cat superposition, however, requires observables which couple states with different $m$. The Heisenberg time evolution of such operators, under $\hat{H}$, involves rapid phase factors $\Exp{\pm i\Omega t}$. Observing quantum interference between the superposed branches of the highly entangled state $\ket{\Psi}$ thus requires measurements that can resolve the short time scale $1/\Omega$. Inspired by the daemon $\hat{H}$ itself, one might perhaps evade this requirement by including a factor $\Exp{\pm ik\hat{Q}}$ in observables; but this would require spatial resolution on the short scale $1/k$. Observing quantum interference between the daemon's Schr\"odinger's Cat branches therefore requires high resolution in either space or time. As far as coarse-grained measurements that lack such resolution are concerned, the actual unitary evolution of the system is indistinguishable from the probabilistic evolution in which the daemon has a random chance, given in the limit $l\gamma \ll\hbar k^{2}/M$ by $\pr_{m}$ as defined in \eqref{prm}, of spontaneously stalling instead of performing its next quantum of work. In other words, as far as any slow observations are concerned, our slow sector effectively loses information through its coupling to the fast sector, and acquires a probabilistic character. Entanglement with fast degrees of freedom is equivalent to decoherence for slow observables.

This is of course the same basic phenomenon to which decoherence, and irreversibility in general, are usually ascribed in macroscopic systems. The small quantum daemon system provides a model in which this phenomenon is rigorously demonstrable. Hamiltonian mechanics means that there is no energy exchange without information exchange, and so steady downconversion implies steadily growing decoherence of the slow sector. In the simple case of a quantum daemon, we can quantify this effect precisely.

If we consider the coarse-grained description of the system which can reproduce measurement outcomes for slow observables only, we must define the reduced density matrix given by tracing over the fast sector,
\begin{equation}
\hat{\rho}_{s}(t) = \sum_{m=-l}^{l}\langle m|_{f}\,|\Psi(t)\rangle\langle\Psi(t)|\,|m\rangle_{f}\;.
\end{equation}
If the total quantum state is pure, the von Neumann entropy of this mixed state is exactly equal to that of the complementary mixed state obtained by tracing over $P$ to obtain a reduced density operator in the $(2l+1)$-dimensional Hilbert space of the fast sector,
\begin{equation}
\rho_{mn}(t) = \int\!dP\,\langle m|_{f}\langle P|_{s}\,|\Psi(t)\rangle\langle\Psi(t)|\,|P\rangle_{s}|n\rangle_{f}\;.
\end{equation}

In the case shown in Figs.~5 and 6, where the momentum width is $D=20\hbar k$, $\rho_{mn}\propto \exp[-100(m-n)^{2}]$, so we can treat $\rho_{mn}$ as diagonal, $\rho_{mn}\doteq R_{m}\delta_{mn}$. Inserting the $P$-dependent time offset and then rescaling $P$ to the dimensionless integration variable $\xi$ lets us further identify
\begin{equation}\label{Rmdef}
R_{m}(t) = \frac{1}{\sqrt{\pi}}\int\!d\xi\,e^{-\xi^{2}}|\Phi_{m}(t-\frac{15 \hbar k}{2Mg}+\frac{\hbar k \xi}{20 Mg})|^{2}
\end{equation}
where $\Phi_{m}(t)$ is the $P$-independent solution to the Schr\"odinger equation
\begin{equation}
i\dot{\Phi}_{m}= \frac{\hbar k^{2}}{2M}\sum_{n=-l}^{l}\left[h_{m}(t)\delta_{mn}-{w}_{mn}\right]\Phi_{n}
\end{equation}
with initial condition $\Phi_{m}(-15\hbar k/(2Mg))=\delta_{ml}$. 

We can therefore straightforwardly calculate the von Neumann decoherence entropy, as a measure of the growth of slow-fast entanglement and thereby of information becoming effectively hidden from the slow sector:
\begin{equation}\label{E:entropy}
S(t) = -\sum_{m=-l}^{l}R_{m}(t)\ln R_{m}(t)\;.
\end{equation}
The result is shown in Fig.~9. If we naively interpreted the Landau-Zener probabilities of \eqref{prm} as applying to abrupt transitions at each avoided crossing, we might expect a step-like growth in decoherence entropy. While step-like entropy growth is indeed possible for a quantum daemon like the one we have analysed, the successive avoided crossings only become so completely isolated from each other for considerably higher values of the parameter $\tilde{\gamma}$. The rougher curve of $S(t)$ shown in Fig.~9 is a measure of the inaccuracy of the naive interpretation of Landau-Zener transitions as instantaneous. The fact that the exact $S(t)$ nonetheless follows the step growth so closely, however, confirms the validity of the Landau-Zener theory for this class of systems. And in particular it confirms that the work done on the weight by the daemon is necessarily accompanied by increasing entropy of the slow subsystem. 

A similar conclusion was recently drawn from Landau-Zener evolution \cite{Barra} in a model for dissipation in a small quantum system that is externally driven while coupled to a macroscopic reservoir. The model consisted of a single fermionic mode with a time-dependent frequency, coupled to a large but finite set of fermionic modes with a discrete set of fixed frequencies. The evolution of the total system in that case was unitary, with the entropy introduced by tracing over part of the system; our entropy is in this respect the same. We justify the tracing out of our fast sector on grounds of spacetime resolution, however, rather than simply on the designation of part of the total system as a reservoir. Our complete system is moreover small, with no large number of modes representing a macroscopic reservoir, and strictly closed, without any time-dependent external parameter to do work on the total system from outside. It is interesting to note how the effective growth of entropy can be derived in both scenarios from Landau-Zener transitions, but our scenario of a quantum Hamiltonian daemon is quite different from dissipation in a driven open system.

\begin{figure}[htb]
	\centering
	\includegraphics[width=.475\textwidth]{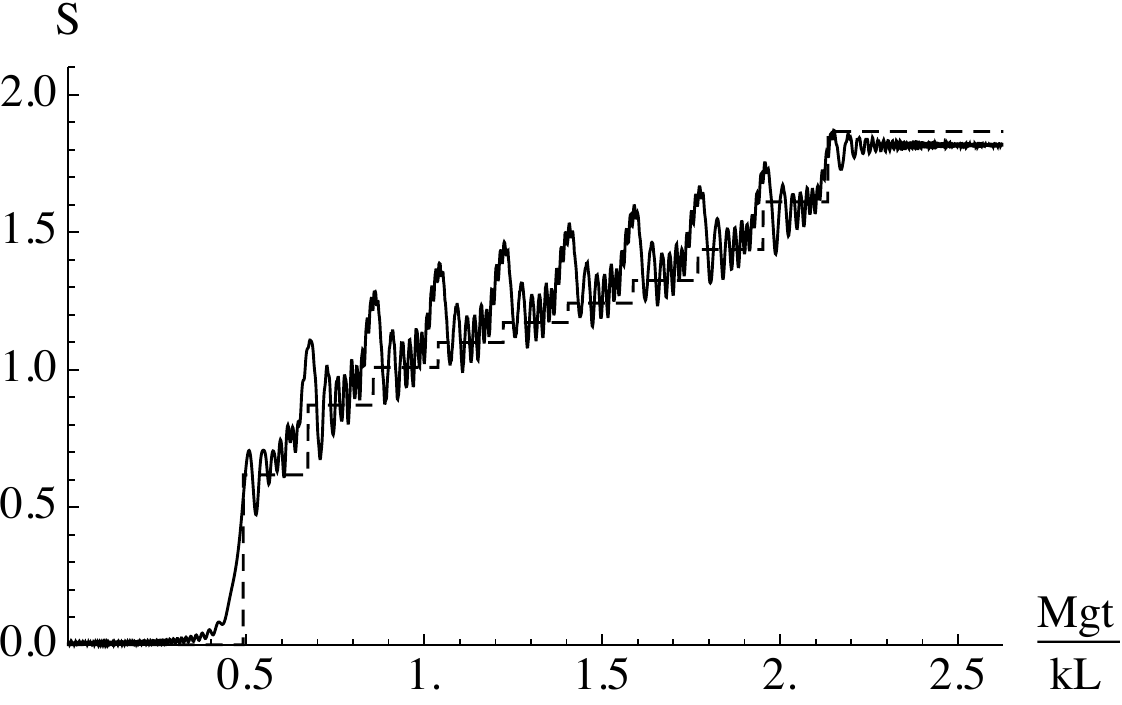}
	\caption{\label{F:entropy} Black curve: The slow-sector von Neumann entropy \eqref{E:entropy} for the exact evolution shown in Figs.~\ref{F:positionSpaceDist} and \ref{F:momentumSpaceDist}. Dashed steps: The approximate entropy obtained by assuming instantaneous Landau-Zener transitions with probabilities given by \eqref{prm}. The rougher growth of the exact curve reflects the inaccuracy of the idealization of isolated avoided crossings, but otherwise the close agreement of the steps with the exact curve confirms the accuracy of the Landau-Zener theory.}
\end{figure}

\subsection{Microthermodynamics}
The paradigmatic system of thermodynamics is the heat engine, which is defined to run as an open system connected to reservoirs, which are by definition macroscopic. Extensions of thermodynamics to microscopic scales have hitherto been considered within this paradigm, and have therefore involved small quantum systems coupled to macroscopic baths \cite{astumian,kieu,hanggi,kim,blickle} and/or time-dependent control parameters \cite{fialko}. It is realistic, however, to take a larger view of heat engines and recognize that the engine's heat reservoir is merely a part of a larger dynamical system. Most practical heat engines are really combustion engines, at least in some sense; the heat bath is simply a means of steadily transferring energy from some form of fuel into some form of work. 

Having taken this larger view of engines, it is paradoxically easier to shrink the engine concept further, and consider fully microscopic analogs to combustion engines, which do not include any macroscopic components at all. The sense in which such systems resemble macroscopic engines can then no longer have anything to do with heat as defined in terms of reservoirs; it can lie, however, in the common achievement of downconversion. The main practical reason why thermodynamics is important, after all, is that engines can extract large amounts of work from small amounts of fuel. The fundamental reason why fuel stores energy so densely is that its degrees of freedom have high dynamical frequencies. Converting this rich source of energy into work is steady downconversion.

The perspective of downconversion thus suggests a certain continuity between the mechanical limitations on Hamiltonian daemons and the thermodynamical limitations on heat engines. This raises the hypothesis that thermodynamics may not actually `emerge' from mechanics in the macroscopic limit, but rather persist into the macroscopic limit, in somewhat modified form, from roots that are already present in the mechanics of simple closed systems---if they exhibit steady downconversion. Since this is a pragmatic perspective, based on the task of performing work with compactly stored energy, the as yet hypothetical subject of microthermodynamics may even one day have practical worth.

\subsection{Decoherence}
If any system is to be practically useful, however, then even if it does not \textit{require} any macroscopic reservoirs, it must be able to operate while in contact with a real environment. The generalization of Hamiltonian daemons to open systems is therefore an important direction for future research. As argued in Ref.~\cite{Us_PRE}, the adiabatic nature of steady downconversion in classical daemons suggests that they may be robust against even moderately strong dissipation and noise. For quantum daemons, however, there is an additional challenge from environmental influence: quantum decoherence \cite{Zurek}.

The quantum daemon that we have studied in this paper operates by unitary time evolution, which preserves quantum superpositions of states. The full many-branched Schr\"odinger's Cat superposition which emerges at the end of the daemon's downconversion phase is sure to decohere into a probabilistic mixture in the presence of any kind of environmental monitoring of the weight's position \cite{Zurek}. It is not clear, however, whether this environmentally induced decoherence will do anything more than enforce the effective decoherence that we have already assumed here as due to limited resolution in observation. 

The only quantum coherence that the extreme quantum daemon actually needs in order to operate is within the two-dimensional Hilbert subspace of each avoided crossing---and this coherence only needs to be maintained to some sufficient degree, over the potentially brief non-adiabatic interval of the Landau-Zener transition. During this interval, the wave packet of the weight may just barely be forking, and realistic environments may be insensitive to the slight distinctions between the two branches at these points. Even environmental couplings which very rapidly decohere true Schr\"odinger's Cat superpositions of macroscopically distinct states may induce only very mild decoherence within a small `quantum halo' subspace of states which are orthogonal in Hilbert space, but negligibly different in their effects on the evolution of the environment \cite{halo}. 

Indeed, the identification and engineering of decoherence-free subspaces has become a significant topic because of its potential usefulness for quantum information technology \cite{Lidar}. On the other hand, a perturbing environment might even make a quantum Hamiltonian daemon operate \textit{more} efficiently, if `slow' environmental monitoring effectively projected the system into adiabatic energy eigenstates \cite{slow1,slow2}, lowering the probability of diabatic stalling below the Landau-Zener limit. Studies of Hamiltonian daemons as open quantum systems are therefore important, but there is no reason to expect that decoherence will be an insuperable barrier to practical realization of quantum daemons.

\begin{acknowledgements}
%\section{Acknowledgements}
JRA gratefully acknowledges valuable discussions with S.~Wimberger, H.~Horner, H.~Ott and A.~Widera. LG acknowledges funding from the German Federal Excellence Initiative (DFG/GSC 266). Both authors thank L.~Thesing, H.J.~Korsch, and two anonymous referees for insightful criticisms of earlier versions of this paper.
\end{acknowledgements}

\begin{appendix}
\section{Relation of $H$ to the model of Ref.~\cite{Us_PRE}}
In our previous work we introduced the Hamiltonian 
\begin{align}
	H_{1} &= \frac{P^{2}}{2M}+\frac{M\nu^{2}}{2}Q^{2}+\frac{p_{+}^{2}+p_{-}^{2}}{2m}+\frac{m}{2}(\Omega_{+}^{2}q_{+}^{2}+\Omega_{-}^{2}q_{-}^{2})\nonumber\\
&-Kq_{+}q_{-}\cos(k Q)\;,
\end{align}
where the coordinates $P$ and $Q$ described the momentum and vertical position of the weight, the coordinates $p_\pm$ and $q_\pm$ described the momenta and positions of two fast harmonic oscillators with frequencies $\Omega_\pm$ respectively and equal masses $m$, and $K$ was a coupling strength. We then introduced new canonical variables $(\tau,U)$ and $(\alpha,A)$ to replace $(q_{\pm},p_{\pm})$:
\begin{eqnarray}\label{AA}
q_{\pm}&=&\frac{\sqrt{\pm 2(U-\Omega_{\mp}A)}}{\sqrt{m\Omega\Omega_{\pm}}}\cos(\Omega_{\pm}\tau+\alpha)\nonumber\\
p_{\pm}&=&-\frac{\sqrt{\pm 2m(U-\Omega_{\mp}A)\Omega_{\pm}}}{\sqrt{\Omega}}\sin(\Omega_{\pm}\tau+\alpha)\;,\\
\Omega &=&\Omega_+-\Omega_-\;.
\end{eqnarray}
(Note that this canonical phase space coordinate $\tau$ from Ref.~\cite{Us_PRE} was unrelated to the dimensionless time $\tau$ in our main text here.)

Using these new variables and discarding small terms which are not resonant anywhere near $\dot Q\doteq \Omega/k$ (see \cite{Us_PRE} for details), we cast $H_1$ into the form 
\begin{eqnarray}\label{H2}
H_{2} &=& \frac{P^{2}}{2M}+\frac{M\nu^{2}}{2}Q^{2} + U\\
&& - \kappa\sqrt{(\Omega_{+}A-U)(U-\Omega_{-}A)}\cos[k(Q-v_{c}\tau)] \nonumber\\
\kappa &=&\frac{K}{2m\Omega\sqrt{\Omega_{+}\Omega_{-}}}\;.\nonumber
\end{eqnarray}
Finally we argued that the potential $M\nu^2/2 Q^2$ would only vary slightly over one period of the rapid oscillations of the downconversion phase, and so we could understand the dynamics adequately by replacing any sufficiently smooth potential with an instantaneously linear potential. This yielded the Hamiltonian
\begin{eqnarray}\label{H3}
	H_{3}&=&\frac{P^{2}}{2M} + M g Q + U \\
&& - \kappa\sqrt{(\Omega_{+}A-U)(U-\Omega_{-}A)}\cos[k(Q-v_{c}\tau)]\;. \nonumber
\end{eqnarray} 
for $g=\nu^2 Q(t)$. If we now consider the transformation
\begin{align}
	L_z&=\frac{U}{\Omega}-\frac{\Omega_++\Omega_-}{2\Omega}A\\
	\phi&=\Omega\tau
\end{align}
which conserves the form of the Poisson brackets involving coordinates $P$, $Q$, $U$, $\tau$, we arrive at the Hamiltonian
\begin{align}
	H_4=&\frac{P^2}{2M}+M g Q+\Omega L_z+\frac{\Omega_++\Omega_-}{2}A\nonumber\\&-\kappa\Omega\sqrt{A^2/4-L_z^2}\cos(kQ-\phi).
\end{align}
Since $A$ is a constant of the motion set by initial conditions we can define $L=A/2$, and identify $H_4$ with $H$ in \eqref{HC} by disregarding the constant energy $L(\Omega_++\Omega_-)$, defining $\gamma=\kappa\Omega$ and expressing the coupling in terms of functions $L_{x,y}$, as defined in the main text, via addition theorems for the trigonometric functions.

\section{Classical reduction from $H$ to $H_\text{eff}$}
As in \cite{Us_PRE}, we make use of the fact that the canonical equations of motion under Hamiltonian \eqref{HC} keep the quantity
\begin{equation}
	J=P+Mg t+k L_z
\end{equation}
exactly constant---even though it is explicitly time-dependent. We exploit this feature by performing the canonical transformation \begin{align}\label{E:trafoReducedCl}
	P&\rightarrow J & Q&\rightarrow Q\\
	L_z&\rightarrow L_z & \varphi&\rightarrow\phi=\phi-kQ,
\end{align}
which is a time-dependent canonical transformation because $J$ is explicitly dependent on $t$ (even though its value remains constant under time evolution).
By inserting \eqref{E:trafoReducedCl} into \eqref{HC} with the correct additional term for the time-dependence of the transformation itself \cite{Goldstein} we find the new Hamiltonian which is equivalent to $H$,
\begin{align}
	H'=&\frac{k^2L_z^2}{2M}-kgL_z\left(t-\frac{J}{Mg}+\frac{\Omega}{kg}\right)\nonumber\\
	&-\gamma\sqrt{L^2-L_z^2}\cos(\phi),
\end{align}
after disregarding a time-dependent energy shift \mbox{$(J-M gt)^2/2M$} and inserting the definitions of the functions $L_{x,y}$ in \eqref{HC}. Up to a possible constant shift in the origin of time we have thus recovered Hamiltonian \eqref{HCred}.

\section{Separatrix area}
The area enclosed by the separatrix can be determined by noting that the upper and lower borders of the separatrix must meet at $\phi=\pm\pi$. This implies that the instantaneous separatrix must be such that
\begin{equation}
\frac{k^{2}}{2M}[L_{z}-gt/k]^{2}+\gamma\sqrt{L^{2}-L_{z}^{2}}=E_{0}(t)
\end{equation}
has a double root $L_{z}=\bar{L}_{z}(t)$. 

Since our concern is with small separatrices, we consider the limit of small $\gamma$. This determines that the latitude of the instantaneous unstable fixed point must be $\bar{L}_{z}(t)=Mgt/k+\mathcal{O}(\gamma)$ and that $E_{0}(t)=\gamma\sqrt{L^{2}-\bar{L}_{z}^{2}(t)}+\mathcal{O}(\gamma^{2})$. This then further determines that the upper and lower borders ${L}_{z\pm}(t)$ of the separatrix at $\phi=0$ must be given by
\begin{equation}
\frac{k^{2}}{2M}[{L}_{z\pm}-\bar{L}_{z}]^{2} - \gamma\sqrt{L^{2}-{L}_{z\pm}^{2}} (\cos\phi+1)= 0
\end{equation}
At least for small enough $\gamma$ we can therefore approximate $L_{z\pm}=\bar{L}_{z}+\mathcal{O}(\sqrt{\gamma})$ and so estimate self-consistently
\begin{equation}
{L}_{z\pm}-\bar{L}_{z} = \pm \sqrt{\frac{4M\gamma\sqrt{L^{2}-\bar{L}_{z}^{2}(t)}}{k^{2}}}\cos(\phi/2).
\end{equation}
Integrating along the upper and lower branches $L_{z\pm}$ with boundaries $-\pi<\phi<\pi$ then yields the estimate of separatrix phase space area as
\begin{equation}
S_\text{sep}\lesssim 16\sqrt{\frac{M\gamma L}{k^2}}\;,%\approx 4\pi L\sqrt{2\tilde M\tilde\gamma}
\end{equation}
as stated in the text.
\end{appendix}

\end{document}